\documentclass{aa}
\usepackage{natbib}
\bibpunct{(}{)}{;}{a}{}{,} 
\usepackage{graphicx}
\usepackage{hyperref}
\usepackage{xcolor}
\usepackage{txfonts}

\usepackage{amsmath}	
\usepackage{longtable}

\makeatletter
\renewcommand*\aa@pageof{, page \thepage{} of \pageref*{LastPage}}
\makeatother

\begin{document}

\title{The story of SN~2021aatd -- a peculiar 1987A-like supernova with an early-phase luminosity excess}

 \author{T. Szalai\inst{1,2,3} \and
        R. K\"onyves-T\'oth\inst{4,5} \and
        A. P. Nagy\inst{1,2} \and 
        D. Hiramatsu\inst{7,8} \and 
        I. Arcavi\inst{9,10} \and 
        A. Bostroem\inst{11} \and 
        D.~A. Howell\inst{12,13} \and 
        J. Farah\inst{12,13} \and
        C. McCully \inst{12,13} \and 
        M. Newsome\inst{12,13} \and 
        E. Padilla Gonzalez\inst{12,13} \and  
        C. Pellegrino\inst{12,13} \and 
        G. Terreran\inst{12,13} \and 
        E. Berger\inst{7,8} \and 
        P. Blanchard\inst{14} \and 
        S. Gomez\inst{15} \and 
        P. Sz\'ekely\inst{1,2} \and 
        D. B{\'a}nhidi\inst{16} \and 
        I. B. B{\'i}r{\'o} \inst{2,16} \and
        I. Cs{\'a}nyi\inst{16} \and 
        A. P{\'a}l\inst{4,5} \and
        J. Rho\inst{17,18} \and
        J. Vink{\'o}\inst{4,5,1,19}
          }
     
   \institute{Department of Experimental Physics, Institute of Physics, University of Szeged, D{\'o}m t{\'e}r 9, 6720 Szeged, Hungary \\ 
              \email{szaszi@titan.physx.u-szeged.hu}
        \and
        HUN-REN--SZTE Stellar Astrophysics Research Group, Szegedi {\'u}t, Kt. 766, 6500 Baja, Hungary 
        \and
        MTA-ELTE Lend\"ulet "Momentum" Milky Way Research Group, Szent Imre H. st. 112, 9700 Szombathely, Hungary 
        \and
        HUN-REN Research Centre for Astronomy and Earth Sciences, Konkoly Observatory, Konkoly Th. M. {\'u}t 15-17., 1121 Budapest, Hungary 
        \and 
        CSFK, MTA Centre of Excellence, Konkoly Thege Mikl{\'o}s {\'u}t 15-17, 1121 Budapest, Hungary 
        \and 
        ELTE E{\"o}tv{\"o}s Lor{\'a}nd University, Gothard Astrophysical Observatory, 9400 Szombathely, Hungary 
        \and
        Center for Astrophysics | Harvard \& Smithsonian, 60 Garden Street, Cambridge, MA 02138-1516, USA 
        \and
        The NSF AI Institute for Artificial Intelligence and Fundamental Interactions 
        \and
        The School of Physics and Astronomy, Tel Aviv University, Tel Aviv 6997801, Israel 
        \and
        CIFAR Azrieli Global Scholars Program, CIFAR, Toronto, Canada 
        \and
        Department of Astronomy, University of Washington, 3910 15th Avenue NE, Seattle, WA 98195-0002, USA 
        \and
        Las Cumbres Observatory, 6740 Cortona Drive, Suite 102, Goleta, CA 93117-5575, USA 
        \and
        Department of Physics, University of California, Santa Barbara, CA 93106-9530, USA 
        \and
        Center for Interdisciplinary Exploration and Research in Astrophysics and Department of Physics and Astronomy, Northwestern University, 1800 Sherman Ave., 8th Floor, Evanston, IL 60201, USA 
        \and
        Space Telescope Science Institute, 3700 San Martin Dr, Baltimore, MD 21218, USA 
        \and
        Baja Astronomical Observatory of University of Szeged, Szegedi {\'u}t, Kt. 766, 6500 Baja, Hungary 
        \and 
        SETI Institute, 339 N. Bernardo Ave., Ste. 200, Mountain View, CA 94043, USA 
        \and
        Department of Physics and Astronomy, Seoul National University, Gwanak-ro 1, Gwanak-gu, Seoul, 08826, South Korea 
        \and
        ELTE E{\"o}tv{\"o}s Lor{\'a}nd University, Institute of Physics and Astronomy, P{\'a}zm{\'a}ny P{\'e}ter s{\'e}t{\'a}ny 1/A, 1117 Budapest, Hungary 
        }

\date{Accepted XXX. Received YYY; in original form ZZZ}

  \abstract
   {While reviewing the current classification scheme of supernovae (SNe) is quite challenging on its own, there is a growing number of peculiar events that cannot be assigned to any of the main classes. SN 1987A and a handful of similar objects, thought to be explosive outcomes of blue supergiant stars, belong to them: while their spectra closely resemble those of H-rich (IIP) SNe, their light-curve (LC) evolution is very different (reaching their primary LC peak in almost 3 months after a short and rapid rise, without any plateaus). Other exciting phenomena are SNe with long-lasting LCs; some of them, like the famous iPTF14hls, have shown multiple LC peaks and an extremely slow spectral evolution.}
   {Here we present the detailed photometric and spectroscopic analysis of SN 2021aatd, a peculiar Type II explosion: while its early-time evolution resembles that of the slowly evolving, double-peaked SN~2020faa (however, at a lower luminosity scale), after $\sim$40 days, its LC shape becomes similar to that of SN~1987A-like explosions.}
   {Beyond comparing LCs, color curves, and spectra of SN~2021aatd to that of SNe 2020faa, 1987A, and of other objects, we compare the observed spectra with our own {\tt SYN++} models and with the outputs of published radiative transfer models. We also carry out a detailed modeling of the pseudo-bolometric LCs of SNe 2021aatd and 1987A with a self-developed semi-analytical code, assuming a two-component (`core' + `shell') ejecta and involving the rotational energy of a newborn magnetar in addition to radioactive decay.}
   {We find that both the photometric and spectroscopic evolution of SN~2021aatd can be well described with the explosion of a $\sim$15 $M_\odot$ blue supergiant star. Nevertheless, SN~2021aatd shows higher temperatures and weaker \ion{Na}{i} D and \ion{Ba}{ii} 6142 \AA\ lines than SN~1987A, which is reminiscent of rather to IIP-like atmospheres. With the applied two-component ejecta model (counting with both decay and magnetar energy), we can successfully describe the bolometric LC of SN~2021aatd, including the first $\sim$40-day long phase showing an excess compared to 87A-like SNe but being strikingly similar to that of the long-lived SN~2020faa. Nevertheless, finding a unified model that also explains the LCs of more luminous events (like SN~2020faa) is still a matter of concern.}
   {}


   \maketitle
%

\section{Introduction}

Type II supernovae (SNe), the cataclysmic endings of
H-rich massive stars, constitute a significant fraction of  of all detected SN explosions \citep[reaching $\sim$35-40\% in galaxies closer than $z \sim0.05$, see e.g.][]{Perley_2020}. Based on their spectral and light-curve (LC) properties, which seem to strongly connect to the rate of lost H envelopes prior to explosion, Type II SNe are divided into various subgroups.

Progenitors of Type IIP events, showing a plateau in their LC, are thought to be red supergiant (RSG) stars with most of their H envelope preserved; while Type IIL SNe show a linear
decline from peak magnitude, probably due to the lower amount of H that could be ionized and recombined after explosion \citep[see e.g.][]{Patat_1994,Hiramatsu_2021}. However, despite some obvious photometric and spectral differences, there are doubts on whether Type IIL and IIP SNe truly originate from different types of progenitors \citep[see, e.g.,][]{Anderson_2014,Valenti_2016}.

A further kind of such explosions, Type IIb ones, constitute an intermediate subgroup between Type II and the H-poor,
stripped envelope Type Ib/c SNe; they can be recognized due to the prominent He features that appear in their spectra after a few days, and the double-peaked shape of their LC (note, however that the first peak is not always detectable). Moreover, we also define Type IIn SN explosions showing
narrow lines in their spectra, indicating interaction between the SN ejecta and a dense circumstellar medium (CSM), see e.g. \citet{schlegel90,filippenko97}. More recently, extremely bright events, called superluminous (SL)SNe, have also been found; those showing a H-rich spectrum are called SLSNe-II \citep{Gal-Yam_2012,BW_2017}. While their origin is still unclear, it has been already revealed that some of them also show signs of strong CSM interaction (similarly to Type IIn ones), while in other cases, alternatives are necessary to explain the extremely high luminosity of these events.

While it is challenging to handle this classification scheme itself, there is a growing number of peculiar events that cannot be assigned to any of the groups described above. SN~1987A, the closest and best-studied extragalactic SN appeared in the Large Magellanic Cloud (LMC), is a typical example: while its spectra closely resemble to those of SNe IIP, its LC evolution is very different \citep[reaching its primary LC peak in almost 3 months after a short and rapid
rise, without any plateaus, see e.g.][]{Hamuy_1988,Whitelock_1988,Catchpole_1988}. Shortly after the first studies, it was revealed that the progenitor of SN~1987A was a compact blue supergiant (BSG), contradicting the stellar evolutionary models developed until that time \citep[e.g.][]{Woosley_1988,Podsiadlowski_1992}. Since the discovery of SN~1987A, extended efforts have been made to find the reason of exploding BSG stars including low metallicity, strong mixing of atmospheric layers, rapid rotation, binary interaction, or stellar mergers \citep[see a recent study on modeling strategies in][]{Dessart_2019}. 
Up to now, only a handful of similar events have been found; however, their number is also growing thanks to the effective analysis of data obtained during transient sky surveys \citep[see e.g.][]{Arcavi_2012,Gonzalez-Gaitan_2015,Taddia_2016,Sit_2023}.

Nowadays, also thanks to the expanding capacities for long-term follow-up of more and more transient events, researchers have the chance to focus on a similarly exciting phenomenon: the existence of SNe with long-lasting, sometimes multipeaked LCs. A strong motivation to look for such kind of events was the discovery of the luminous, extraordinarily long-lived Type II iPTF14hls that has showed at least five LC peaks and an extremely slow spectral evolution \citep[reaching the nebular phase after more than 600 days,][]{Arcavi_2017,Sollerman_2019}.
Among others, enhanced CSM interaction, magnetar formation, fall-back accretion, common-envelope jets, and pulsational pair-instability SNe have been invoked to explain this unique phenomenon \citep[see][for a review]{Woosley_2018}. During the extended analysis of Zwicky Transient Facility (ZTF) data, \citet{Soraisam_2022} identified almost 40 SNe of various types showing multipeaked LCs, including SN~2020xkx, a true photometric analog of iPTF14hls (unfortunately, without a well-sampled spectroscopic follow-up). Another ZTF object was SN~2020faa, which seemed to be also quite similar to iPTF14hls (at least, during the first $\sim$200 days of its evolution) and has been analyzed in detail by \citet{Yang_2021} and \citet{Salmaso_2023}.

In this work, we present the detailed photometric and spectroscopic analysis of SN~2021aatd, another peculiar Type II explosion. 
While its early-time evolution resembles to that of SN~2020faa and iPTF14hls (however, at a lower luminosity scale), after $\sim$40 days, its LC shape becomes definitely similar to that of SN~1987A-like explosions. 
This paper is organised as follows. 
First, we present our ground-based photometric and spectroscopic
observations in Sect. \ref{sec:obs}. Then, we describe the steps of comparative LC and spectral modeling analysis in Section~\ref{sec:anal}.
Finally, we discuss the results and provide our concluding remarks in Section~\ref{sec:concl}. 

\section{Observations}\label{sec:obs}

\begin{figure*}
    \centering
   \includegraphics[width=0.95\textwidth]{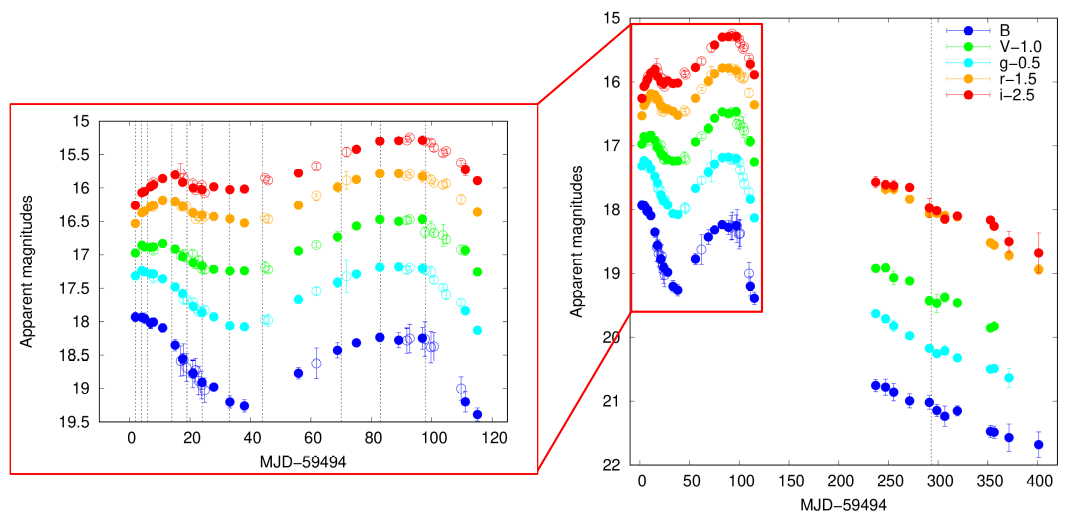}
     \caption{$BVgri$ photometry of SN~2021aatd. Filled and empty circles denote measurements are from LCO sites and from Baja Observatory, respectively. Dotted vertical lines mark the epochs of spectroscopic observations shown in Table \ref{tab:spec}.}
    \label{fig:lc}
\end{figure*}

SN~2021aatd (ATLAS21bjwk) was discovered by the Asteroid Terrestrial-impact Last Alert System \citep[ATLAS,][]{Tonry_2018,Smith_2020atlas} on  2021 October 7.38 UT (MJD 59494.38), at an apparent brightness of m$_o$ = 18.35$\pm$0.10 mag \citep{Chen_2021aatd}, with an epoch of last non-detection on 2021 October 5.51 UT (MJD 59492.51). The SN is offset by 2.1 arcsecs south, 5.7 arcsecs west from its likely host galaxy GALEXASCJ005904.40-001210.0. The first spectroscopic classification was reported by \cite{Wyatt_2021aatd} who found the object being consistent with a young, peculiar Type II SN (showing the best match with the spectrum of SN~1987A). 

The redshift of the host galaxy is given as z = 0.01524$\pm$0.00004 in 
NASA/IPAC Extragalactic Database (NED\footnote{\href{https://ned.ipac.caltech.edu/}{https://ned.ipac.caltech.edu/}}), corresponding to a Hubble-flow distance of d = 62.6$\pm$4.4 Mpc (assuming H$_0$ = 67.8 $\pm$ 4.7 km s$^{-1}$ Mpc$^{-1}$). The Galactic extinction of $E(B-V)=0.025$ was also adopted from NED \citep{SF11}. Because of the lack of early-time high-resolution spectra, we do not have any information on the extinction in the host galaxy; however, based on the position of the SN, host reddening is probably not significant.


\subsection{Photometry}
\label{sec:obs_phot}

\begin{figure}
\centering
\includegraphics[width=\columnwidth]{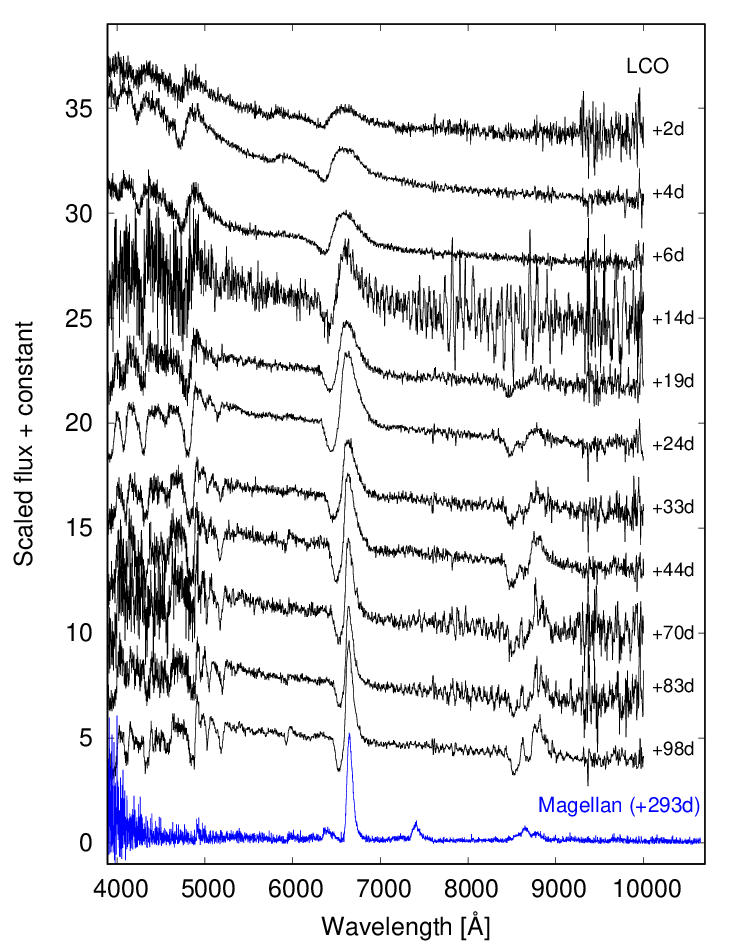}
\caption{Optical spectra of SN~2021aatd obtained from LCO sites (black), and with Magellan (blue).}
\label{fig:sp_all}
\end{figure}

We performed optical Johnson $BV$ and Sloan $gri$ (hereafter $BVgri$) photometry of SN~2021aatd with follow-up observations from Las Cumbres Observatory (LCO) utilizing a world-wide network of telescopes under the Global Supernova Project \citep[GSP,][]{Howell_2019}, and with the 0.8m Ritchey-Chrétien telescope at the Baja Observatory, Hungary \citep{Barna_2023}. 
The median combination and the aperture photometry of LCO data was carried out applying self-written pipelines using the {\tt fitsh} software \citep{pal12}. 
Data obtained at Baja Observatory were processed with standard IRAF\footnote{IRAF is distributed by the National Optical Astronomy Observatories, which are operated by the Association of Universities for Research in Astronomy, Inc., under cooperative agreement with the National Science Foundation.} routines, including basic corrections. Then we co-added three images per filter per night aligned with the {\tt wcsxymatch}, {\tt geomap} and {\tt geotran} tasks. We obtained PSF photometry on the co-added frames using the {\tt daophot} package in IRAF, and image subtraction photometry based on other IRAF tasks like {\tt psfmatch} and {\tt linmatch}, respectively. For the image subtraction we applied a template image taken at a sufficiently late phase, when transient was no longer detectable on our frames. 

The photometric calibration was carried out using
stars from Data Release 1 of Pan-STARRS1 (PS1 DR1) \footnote{\href{https://catalogs.mast.stsci.edu/panstarrs/}{https://catalogs.mast.stsci.edu/panstarrs/}}. 
In order to get reference magnitudes for our $B$- and $V$-band frames, the PS1 magnitudes were transformed into the Johnson $BVRI$ system based on equations and coefficients found in \cite{Tonry12}. Finally, the instrumental magnitudes were transformed into standard $BVgri$ magnitudes by applying a linear color term (using $g-i$) and wavelength-dependent zero points. Since the reference stars fell within a few arc-minutes around the target, no atmospheric extinction correction was necessary. S-corrections were not applied.
$BVgri$ photometry of SN~2021aatd obtained from LCO sites and from Baja Observatory are collected in Tables \ref{tab:phot_data_LCO} and \ref{tab:phot_data_Baja}, respectively; light curves (LCs) are plotted in Fig. \ref{fig:lc}.

\subsection{Spectroscopy}
\label{sec:obs_spec}

\begin{figure*}
    \centering
    \includegraphics[width=0.95\textwidth]{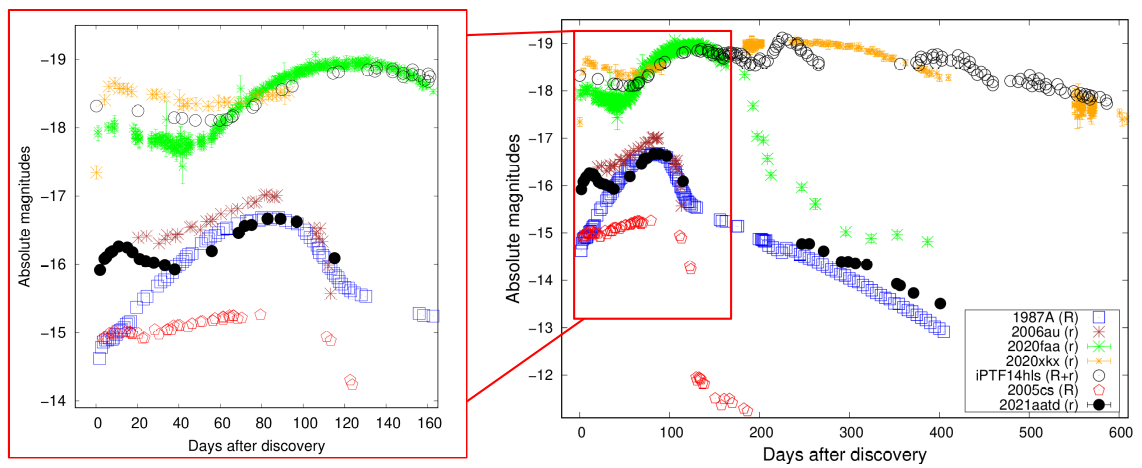}
    \caption{Comparison of the early- and late-time $r$-band photometric evolution of SN~2021aatd to that of iPTF14hls \citep[$r+R$ filters,][]{Arcavi_2017}, SNe 1987A \citep[$R$ filter,][]{Hamuy_1988,Whitelock_1988,Catchpole_1988,Catchpole_1989}, 2005cs \citep{Pastorello_2009}, 2006au \citep{Taddia_2012}, 2020faa \citep{Yang_2021,Salmaso_2023}, and 2020xkx \citep[][downloaded from the ALeRCE ZTF Explorer]{Soraisam_2022}. Note that in the case of SN~2006au and iPTF14hls, there is a large uncertainty regarding the date of explosion; thus, their LCs are shifted to match with the LC maxima of SNe~1987A and 2020faa, respectively \citep[following the method of][]{Taddia_2012,Yang_2021}.}
    \label{fig:lc_comp}
\end{figure*}

\begin{figure}
    \includegraphics[width=\columnwidth]{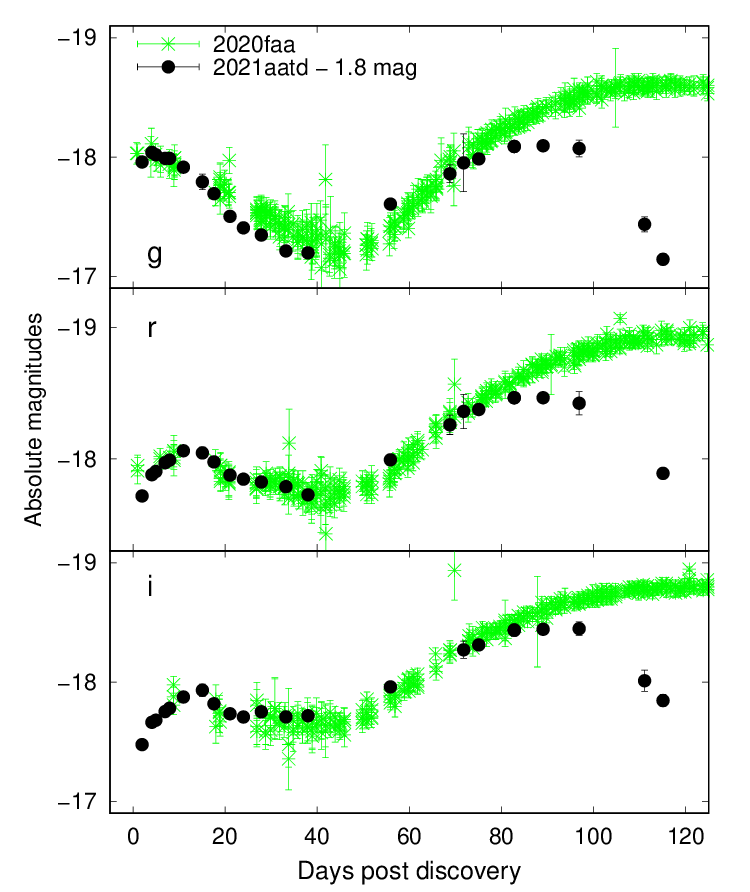}
    \caption{Comparison of early-phase absolute magnitudes of SNe~2020faa and 2021aatd. Latter values are shifted by $-$1.8 mag.}
    \label{fig:20faa_comp}
\end{figure}

\begin{figure*}
    \includegraphics[width=0.5\textwidth]{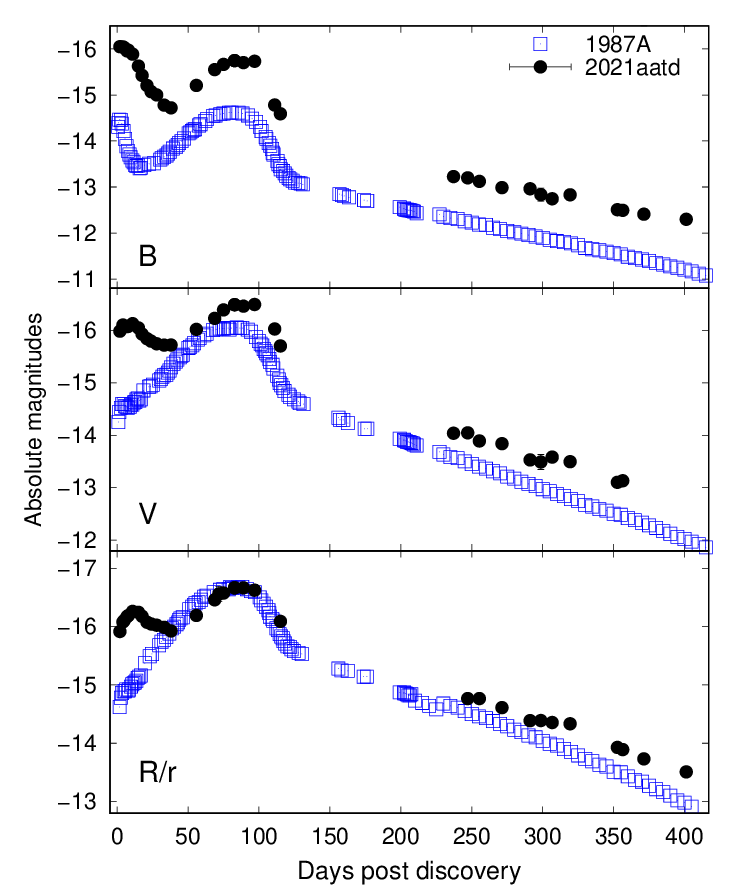}
    \includegraphics[width=0.5\textwidth]{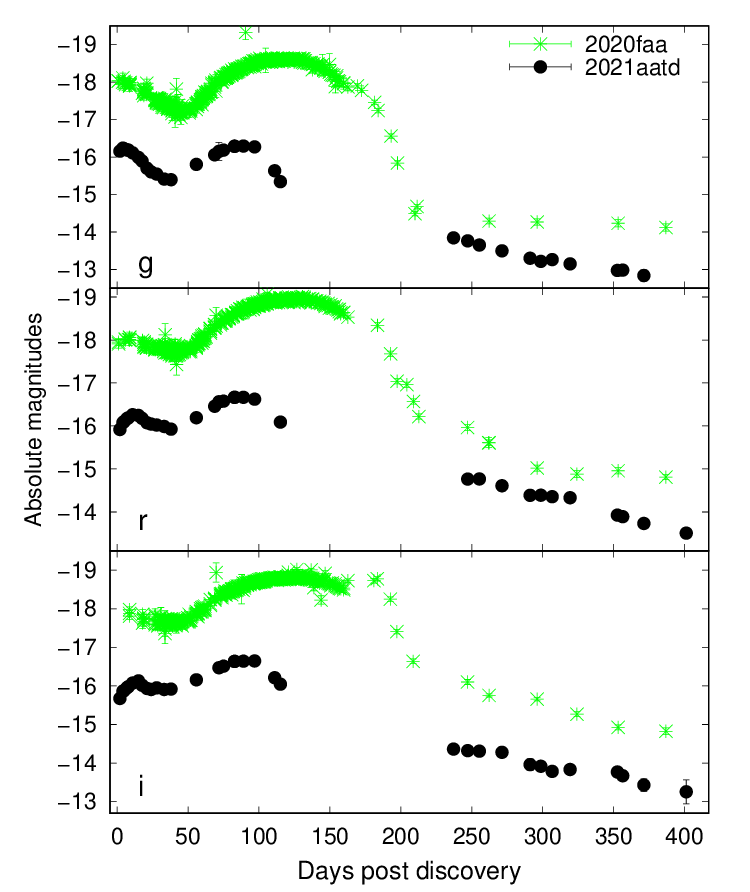}
    \includegraphics[width=0.5\textwidth]{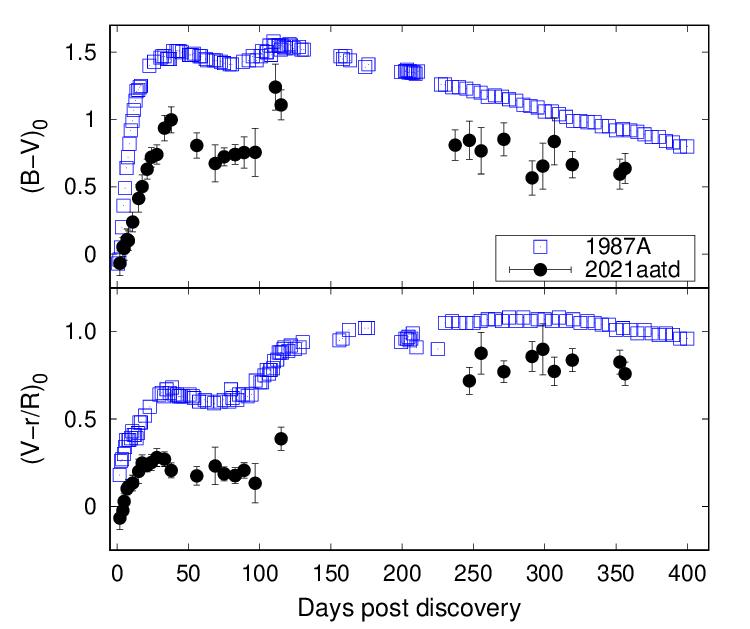}
    \includegraphics[width=0.5\textwidth]{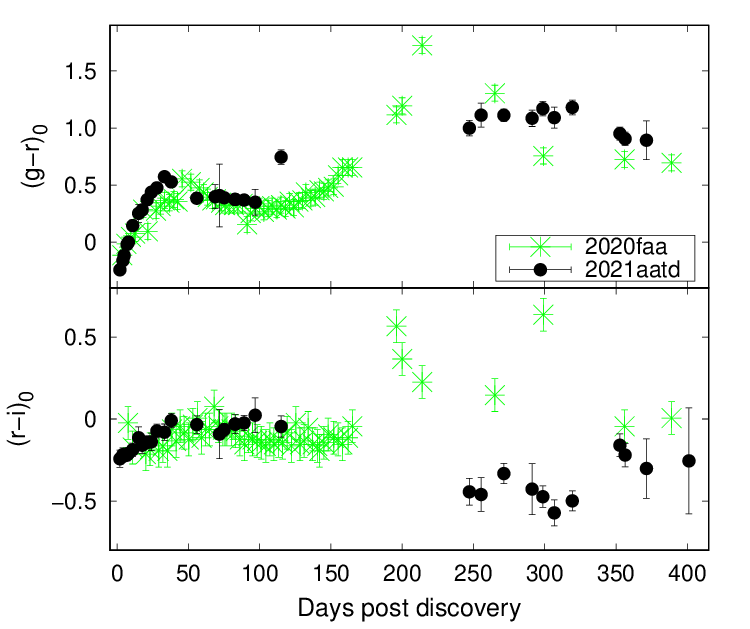}
    \caption{Comparison of the light- and color curves of SN~2021aatd to that of SNe~1987A and 2020faa.}
    \label{fig:color_comp}
\end{figure*}

We obtained 11 sets of optical spectra at the LCO
with the FLOYDS spectrographs mounted on the 2m
Faulkes Telescope North (FTN) at Haleakala (USA) and the identical 2m Faulkes Telescope South (FTS) at Siding Spring (Australia), through the GSP, between 2021 Oct 9 and 2022 January 13. A 2\arcsec-wide slit was placed on the target at the parallactic angle \citep{filippenko82}. We extracted, reduced, and calibrated 1D spectra following standard procedures using the FLOYDS pipeline \citep{Valenti_2014}.

One late-time spectrum was obtained on 2022 July 26 with the Low Dispersion Survey Spectrograph 3 (LDSS-3) on the 6.5m Magellan Clay telescope at Las Campanas Observatory in Chile. Standard reductions were carried out using IRAF, including bias subtraction, flat-fielding, cosmic-ray rejection, local sky subtraction, and extraction of one-dimensional spectra. The slit was aligned along the parallactic angle to minimize differential light losses, and flux calibration was done using a spectrophotometric standard taken that night at similar airmass. 

The log and a summarizing plot of our spectroscopic observations on SN~2021aatd are shown in Table \ref{tab:spec} and in Fig. \ref{fig:sp_all}, respectively.

\begin{table*}
\begin{center}
\caption{Peak times and magnitudes in the LCs of SN~2021aatd}
\label{tab:LC_peaks}
\begin{tabular}{c|ccc|ccc}
\hline
\hline
~ & $T_\textrm{peak,1}$ (MJD) & Peak mag$_1$ & Peak abs. mag$_1$ & $T_\textrm{peak,2}$ (MJD) & Peak mag$_2$ & Peak abs. mag$_2$ \\
\hline
$B$ & 59497.8 $\pm$ 2.2 & 17.94 $\pm$ 0.08 & $-$16.13 $\pm$ 0.17 & 59581.1 $\pm$ 0.8 & 18.25 $\pm$ 0.04 & $-$15.83 $\pm$ 0.16 \\
$V$ & 59503.9 $\pm$ 2.3 & 17.83 $\pm$ 0.04 & $-$16.22 $\pm$ 0.16 & 59581.4 $\pm$ 1.0 & 17.45 $\pm$ 0.05 & $-$16.60 $\pm$ 0.16 \\
$g'$ & 59499.9 $\pm$ 0.6 & 17.76 $\pm$ 0.04 & $-$16.31 $\pm$ 0.16 & 59581.6 $\pm$ 0.7 & 17.67 $\pm$ 0.04 & $-$16.40 $\pm$ 0.16 \\
$r'$ & 59506.4 $\pm$ 1.5 & 17.66 $\pm$ 0.08 & $-$16.38 $\pm$ 0.17 & 59581.6 $\pm$ 0.5 & 17.28 $\pm$ 0.03 & $-$16.76 $\pm$ 0.16 \\
$i'$ & 59508.2 $\pm$ 3.6 & 17.80 $\pm$ 0.06 & $-$16.23 $\pm$ 0.16 & 59583.0 $\pm$ 1.0 & 17.26 $\pm$ 0.06 & $-$16.77 $\pm$ 0.16 \\
\hline
\end{tabular}
\end{center}
\end{table*}

\begin{table*}
\begin{center}
\caption{Basic data of Type II SNe used for comparison}
\label{tab:sne}
\begin{tabular}{lccccc}
\hline
\hline
Name & Date of discovery & $z$ & $D$ & E(B$-$V) & Source \\
~ & (MJD) & ~ & (Mpc) & (mag) & ~ \\
\hline
{\bf SN~2021aatd} & {\bf 59494.4} & {\bf 0.015$^a$} & {\bf 62.6$\pm$4.4$^a$} & {\bf 0.025$^a$} & {\bf This work} \\
SN~1987A & 46849.8 & 0.0009 & 0.051 & 0.180 & \citet{Hamuy_1988} \\
SN~2005cs & 53549.0 & 0.0018 & 7.1$\pm$1.2 & 0.050 & \citet{Pastorello_2009} \\
SN~2006au & 53801.2 & 0.0098 & 46.2$\pm$3.2 & 0.300 & \citet{Taddia_2012} \\
iPTF14hls & 56922.5 & 0.035 & 156 & 0.014 & \citet{Arcavi_2017} \\
SN~2020faa & 58933.1 & 0.041 & 187 & 0.022 & \citet{Yang_2021} \\
SN~2020xkx & 59140.0 & 0.042 & 190 & 0.070 & \citet{Soraisam_2022} \\
\hline
\end{tabular}
\end{center}
\smallskip
{\bf Notes.} $^a$Adopted from NASA/IPAC Extragalactic Database (NED, https://ned.ipac.caltech.edu).
\end{table*}

\section{Analysis and Results}\label{sec:anal}

\subsection{Light-curve \& color evolution}\label{sec:anal_phot}

SN~2021aatd shows a double-peak LC in all filters. we fit single Gaussian functions to the brightness values around the peaks to get proper estimations on the peak magnitudes and epochs (see Table \ref{tab:LC_peaks}). Blue-band LCs peak systematically earlier than redder bands (note that $B$-peak is very close to the epoch of our first observation).
After the initial peak, the SN shows a declining phase up to $\sim$40 days when it starts to brighten again. In all filters, except $B$-band, the SN reaches a second peak magnitude higher, than the previous one. The second peak of the LC is monitored with a higher cadence -- while, considering the mean values of the second peak times, there is a 2-days shift from $B$ to $i$-band maximum, their uncertainty ranges actually overlap.  
After the second peak, the LCs of SN~2021aatd show a rapid decline of about 0.6$-$1.1 mag (from the longest to the shortest wavelength, respectively) in $\sim$20 days before the SN disappeared behind the Sun. Observations continued from $\sim$240 d up to $\sim$400 d after discovery; in this phase, the SN showed a slow linear declining tail.

The late-time LC evolution of SN~2021aatd -- peak brightness reached $\sim$3 months after explosion, followed by a rapid decline and, finally, by a linear declining tail -- resembles quite well to the behaviour of 1987A-like SNe. Nevertheless, neither SN~1987A, nor the few known similar events assumed to be explosions of blue supergiant stars, show the first peak in their LCs \citep[see a recent overview in][]{Xiang_2023}. This suggests that, during the first several weeks of the evolution, a different LC powering mechanism may exist in the case of SN~2021aatd.
We searched objects with a similar early-time behaviour in recently published studies focusing on optically rebrightening SNe. As we show in detail below, the early-phase LC evolution of SN~2021aatd is very similar to that of special Type II SN~2020faa \citep{Yang_2021,Salmaso_2023} -- however, SN~2020faa was a more luminous event. 

In Fig. \ref{fig:lc_comp}, we present the comparison of the early- and late-time $r$-band photometric evolution of SN~2021aatd to those of SN~2020faa, SN~ 1987A \citep[$R$ filter,][]{Hamuy_1988,Whitelock_1988,Catchpole_1988,Catchpole_1989} the 87A-like event SN~2006au \citep{Taddia_2012}, the low-luminosity Type IIP SN~2005cs \citep{Pastorello_2009}, as well as the rebrightening objects iPTF14hls \citep[$r+R$ filters,][]{Arcavi_2017} and SN~2020xkx \citep[][downloaded from the ALeRCE ZTF Explorer\footnote{\href{https://alerce.online/}{https://alerce.online/}}]{Soraisam_2022}. Basic data of the SNe used for generating absolute magnitudes are shown in Table \ref{tab:sne}. 
Note that, in the case of SN~2006au and iPTF14hls, there is a large uncertainty regarding the date of explosion; thus, their LCs are shifted to match with the LC maxima of SNe~1987A and 2020faa, respectively \citep[following the method of][]{Taddia_2012,Yang_2021}.

As it can be seen in Fig. \ref{fig:lc_comp}, 87A-like and rebrightetning SNe (iPTF14hls, 2020faa, 2020xkx) form two distinct luminosity groups. Note, however, that LC of SN~2020faa starts to quickly decline $\sim$200 days after explosion, on the contrary to the other two objects \citep[see][]{Salmaso_2023}. SN~2021aatd, apparently, belongs to the group of 87A-like events. At the same time, the first $\sim$40 days of its LC evolution closely resemble to that of bright, slowly evolving SNe, especially of 2020faa. 
Early-phase LC similarities of SNe 2021aatd and 2020faa can be seen even better in Fig. \ref{fig:20faa_comp}, which shows the comparison of early-phase absolute LCs not only in $r$, but also in $g$ and $i$ filters. To get an almost perfect match, one should shift the LCs of SN~2021aatd by $\sim$1.8 mag.

SN~2006au seems to also show some early-phase flux excess compared to the LC shape of SN~1987A (note that, in general , SN~2006au was a bit more luminous event). Unfortunately, the earliest part of the LC of SN~2006au is probably missing \citep[the discovery of the object occured more than one year since the previous non-detection, see][]{Taddia_2012}; thus, it is difficult to use this object for a detailed comparative analysis. We also omitted iPTF14hls and SN~2020xkx from the further LC analysis, since their late-time evolution is completely different from that of SN~2021aatd. Therefore, we only show a detailed comparison of light and color curves of SN~2021aatd to that of SNe~1987A and 2020faa (based on $B$, $V$, $R/r$, and $g$, $r$, $i$ data, respectively), see Fig. \ref{fig:color_comp}.

SN~2021aatd, as has already been presented in Fig. \ref{fig:20faa_comp}, shows a very similar early-time LC evolution to that of SN~2020faa, but at lower luminosities, which can be also well seen in both $g-r$ and $r-i$ curves (Fig. \ref{fig:color_comp}, bottom right panel). We also discussed above the general differences between the late-time LCs of the two events. Moreover, taking another look at the right panels of of Fig. \ref{fig:color_comp}, the slope of the LCs of SNe 2020faa and 2021aatd seem to be quite similar in $i$ band, again, while this is not the case in the $g$ and $r$ bands \citep[note however, that very late-time $g$- and $r$-band data of SN~2020faa are close to the detections limit, see][]{Salmaso_2023}.
After comparing the light and color curves of SN~2021aatd with that of SN~1987A, we see similar evolution after the first $\sim$50 days. However, SN~2021aatd seems to be a bit bluer in general -- but, note again that we have no information on the host-galaxy extinction in the direction of SN~2021aatd.

\subsection{Spectral analysis}\label{sec:anal_sp}

\begin{figure*}
\centering
\includegraphics[width=.95\columnwidth]{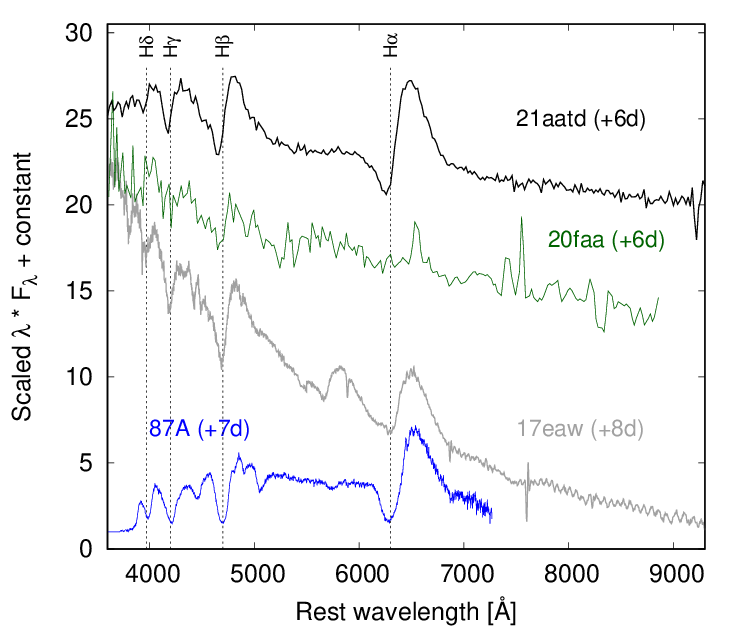}
\includegraphics[width=.95\columnwidth]{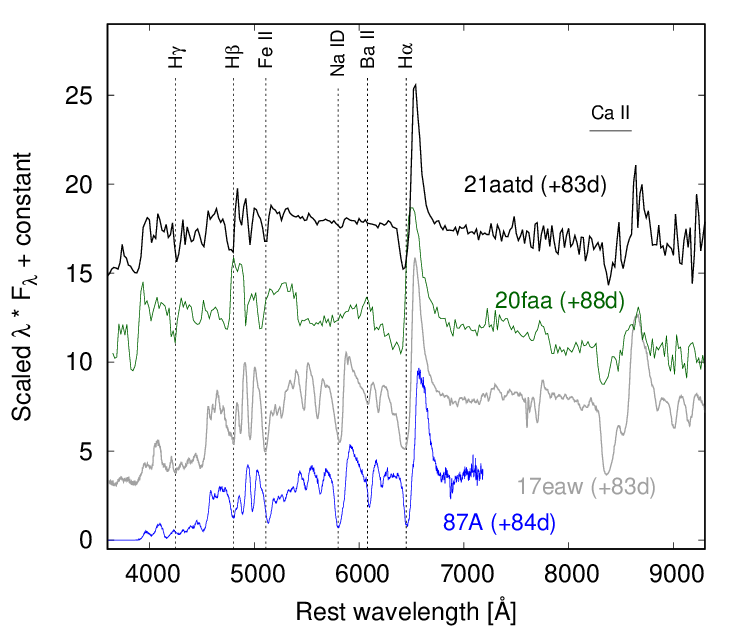}
\caption{{\it Left:} The comparison of the +6d spectrum of SN~2021aatd to that of SNe~2020faa \citep{Yang_2021}, 2017eaw \citep[IIP,][]{szalai19b} and 1987A \citep{Pun_1995} obtained at +6, +8, and +7 days phase, respectively. All the spectra are corrected for redshift and extinction. Spectra of SNe~1987A and 2020faa have been downloaded from WISeREP database. {\it Right:} The same comparison of spectra of the four SNe obtained at +83, +88, +83, and +84 days, respectively.}
\label{fig:sp_comp}
\end{figure*}

After the detailed comparative LC analysis, in Fig. \ref{fig:sp_comp}, we present a set of spectra of SNe 2021aatd, 2020faa, and 1987A obtained at both a very early epoch (6-8 days after explosion) and around the second LC peak of SN~2021aatd (+83-88 days). For a further comparison, we also plot the same-age spectra of SN~2017eaw, one of the best-studied Type IIP explosion.

As can be seen, in accordance with the color evolution shown in Fig. \ref{fig:color_comp}, SNe~2021aatd and 2020faa have basically the same temperature evolution up to the peak brightness, while SN~1987A has a redder spectrum at both epochs. SN~2017eaw shows the bluest spectrum at early time, but, at +83/88 days, its temperature becomes definitely lower than that of SNe 2021aatd and 2020faa. The slow spectral evolution is a well-known characteristic of multiple LC-peaked SNe like iPTF14hls or SN~2020faa \citep{Arcavi_2017,Sollerman_2019,Yang_2021,Salmaso_2023}; up to $\sim$100 days after explosion (when its last ''photospheric'' spectrum was obtained), 2021aatd seems to also belong to this group.
At both early and later epochs, spectra of SN~2021aatd show strong H Balmer lines, similarly to that of SNe IIP and 1987A. At the same time, one can see differences in the cases of \ion{Na}{i} D and \ion{Ba}{ii} 6142 \AA\, two other spectral features definitely worth analyzing during the comparison of Type II SNe. In the case of SNe 1987A and 2017eaw, \ion{Na}{i} D has a deep P Cygni profile at +83/88 days, while SNe 2021aatd and 2020faa shows only weak \ion{Na}{i} D lines. In the spectra of SN~1987A, \ion{Ba}{ii} 6142 \AA\ is also a dominant line, while this is not the case in spectra of other three objects.  It has been described in  several works that there is a diversity in the strength of \ion{Ba}{ii} 6142 \AA\ line among 87A-like SNe, and weakness of this line is rather a characteristic of IIP-like SN atmospheres \citep[see e.g.][]{Takats_2016,Dessart_2019,Xiang_2023}. 

Using our optical spectra -- following the general method also used in the cases of SN~2020faa \citep{Yang_2021,Salmaso_2023} and of 87A-like SNe \citep[e.g.][]{Xiang_2023} --, we determined the H$\alpha$ and \ion{Fe}{ii} 5169\AA\ line velocities for SN~2021aatd. Taking advantage of the adequate signal-to-noise ratio of our spectra, we fit single Gaussian profiles to the regions of the absorption minima of the two lines for calculating the line velocities ($v_{\text{H}\alpha}$ and $v_{\text{FeII}}$). 
Velocity curves of SN~2021aatd, together with that of SNe 1987A, 2020faa, and Type IIP 2012aw \citep[also used for comparison by][]{Salmaso_2023}, are shown in Fig. \ref{fig:sp_vel_comp}. We note that, in the lack of spectrum for SN~2012aw around +80-90 days, we used the data of its spectral twin, SN~2017eaw \citep{szalai19b} for the spectral comparison shown in Fig. \ref{fig:sp_comp}. 

In order to monitor the chemical composition, as well as the evolution of the (photospheric) temperatures and velocities in more detail, we carried out {\tt SYN++} modeling for the spectra of SNe~2021aatd and 2020faa; moreover, we also compared our observational data to the outcomes of models published by \citet{Dessart_2019}.

\subsubsection{{\tt SYN++} modeling of the photospheric spectra}
We applied the parameterized resonance scattering code {\tt SYN++}\footnote{\href{https://c3.lbl.gov/es/}{https://c3.lbl.gov/es/}} \citep{thomas11} for the analysis of 2$-$98d spectra of SN~2021aatd. Note that, basically, the code can be used for analyzing photospheric spectra; however, because of its slow spectral evolution, SN~2021aatd can be still considered in the photospheric phase at +98 days. The code allows to make estimations on some global parameters, such as the photospheric temperature ($T_{\rm ph}$), and the velocity at the photosphere ($v_{\rm ph}$). The contribution of the single ions can be taken into account by setting some local parameters -- the optical depth ($\tau$), the minimum and the maximum velocity of the line forming region ($v_{\rm min}$ and $v_{\rm max}$), the scale height of the optical depth above the photosphere (aux), and the excitation temperature ($T_{\rm exc}$) -- for each identified ion.
Since we did not find any similar analyses for SN~2020faa in the literature, we also involved its photospheric spectra in our modeling process. Moreover, as a consistency check, we also fit {\tt SYN++} models on the +84d spectrum of SN~1987A. 

\begin{figure*}
\centering
\includegraphics[width=.95\columnwidth]{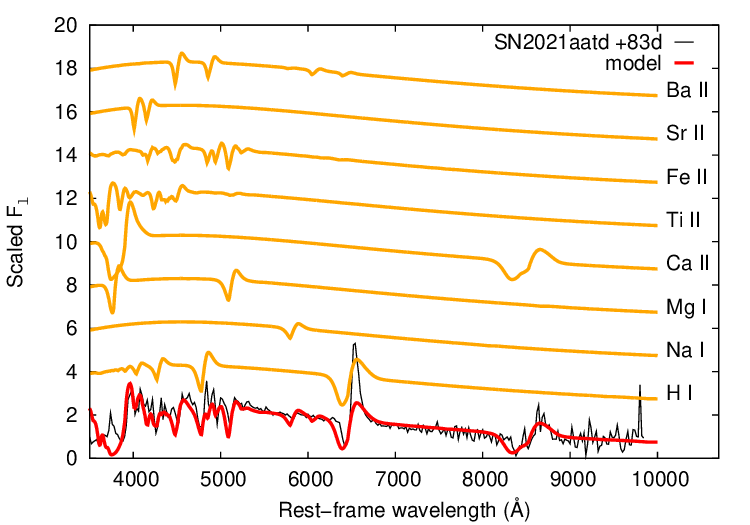}
\includegraphics[width=.95\columnwidth]{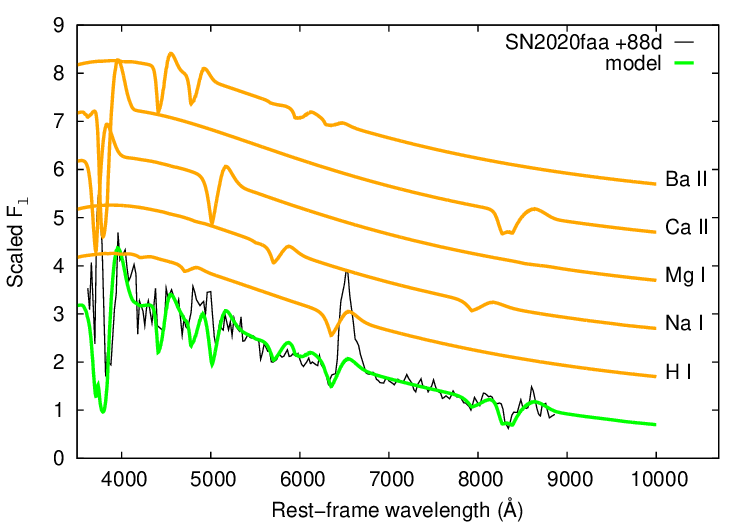}
\includegraphics[width=.95\columnwidth]{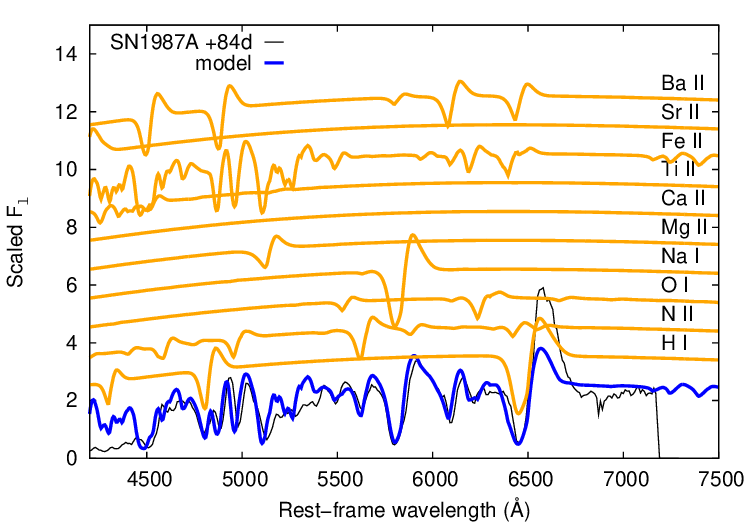}
\caption{Ion contributions from SYN++ models calculated for +83, +88, and +84 days spectra of SNe 2021aatd, 2020faa, and 1987A, respectively.}
\label{fig:sp_comp2}
\end{figure*}

\begin{figure*}
\centering
\includegraphics[width=.95\columnwidth]{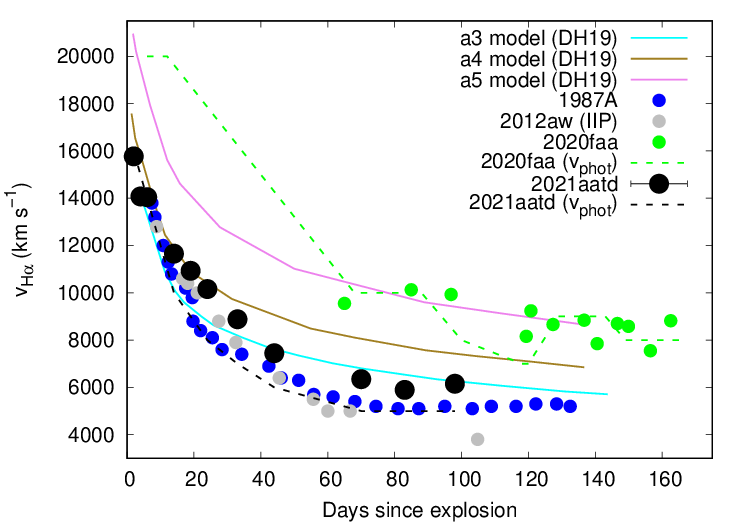}
\includegraphics[width=.95\columnwidth]{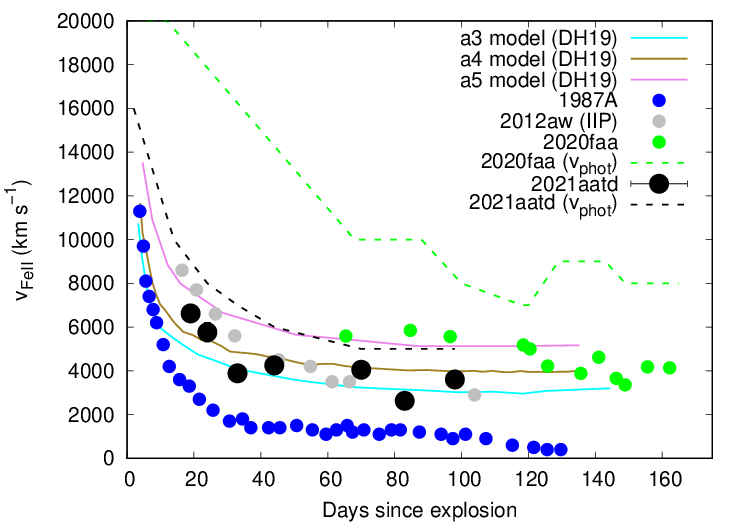}
\includegraphics[width=.95\columnwidth]{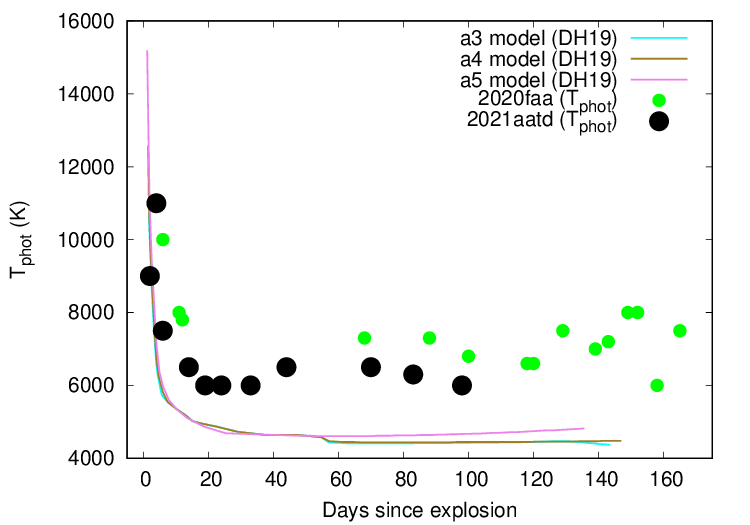}
\includegraphics[width=.95\columnwidth]{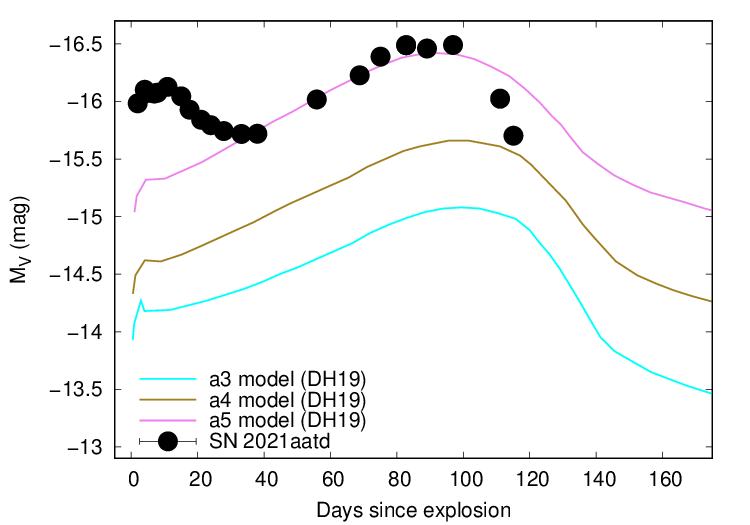}
\caption{Measured H$\alpha$ ({\it top left}) and Fe II 5169 \AA\ ({\it top right}) line velocities of SN 2021aatd, compared to that of SNe 1987A \citep{Xiang_2023}, 2020faa \citep{Yang_2021}, and Type IIP 2012aw \citep{bose2013}. Calculated line velocities from {\it a3}, {\it a4}, and {\it a5} explosion models of \citet{Dessart_2019} (DH19), as well as photospheric velocities of SNe 2021aatd and 2020faa, determined from our {\tt SYN++} models, are also plotted. See text for further details. {\it Bottom left:} Calculated photospheric temperatures ($T_{\rm phot}$) from DH19 {\it a3}, {\it a4}, and {\it a5} models, compared to $T_{\rm phot}$ values of SNe 2021aatd and 2020faa determined from our {\tt SYN++} models.
{\it Bottom right:} Calculated V-band absolute magnitudes ($M_{\rm V}$) from DH19 {\it a3}, {\it a4}, and {\it a5} models, compared to the measured values of SN~2021aatd.}
\label{fig:sp_vel_comp}
\end{figure*}

Fig. \ref{fig:21aatd_model} shows the results of {\tt SYN++} modeling carried out on all the 11 spectra of SN~2021aatd obtained during the first $\sim$3 months after explosion. The observed spectra were corrected for extinction and redshift before plotting. 
The values of the best-fit global and local parameters are collected in Table \ref{tab:local_21aatd}.
At the earliest epochs, \ion{H}{i} and \ion{He}{ii} lines dominate the spectrum of SN~2021aatd, which can be best described with the global parameters of $v_{\rm ph} \sim$18\,000 km s$^{-1}$ and $T_{\rm ph} \sim$6500 K. At later phases (after $\sim$20 days), lines of \ion{Na}{i}, \ion{Mg}{i}, \ion{Mg}{ii}, \ion{Ca}{ii}, \ion{Ti}{ii}, \ion{Sr}{ii}, and \ion{Ba}{ii} emerge as well and give a significant contribution to the spectrum formation. By +98 days (the last observed epoch of the photospheric phase), $v_{\rm ph}$ decreases to 5000 km s$^{-1}$, while $T_{\rm ph}$ shows a flat, nearly constant evolution around 5000 K.
We note that there is a known inconsistency during the parallel fitting of both the absorption and emission components of the strong H$\alpha$ line profile -- this is probably due to the basic assumption of the {\tt SYN++} code of using local thermodynamic equilibrium (LTE) for calculating level populations \citep[see e.g.][]{szalai16}.

For comparison, we also modeled the spectra of SN~2020faa \citep[obtained at 13 different epochs from +6 to +165 days, published by][]{Yang_2021} with {\tt SYN++}. 
Plots and parameters are shown in Fig. \ref{fig:20faa_model} and in Table \ref{tab:local_20faa}, respectively, in the same way as in the case of SN~2021aatd. 
Both the list of identified ions and the overall chemical evolution of SN~2020faa (at least, during the first 3 months) shows a great resemblance with that of SN~2021aatd -- even though SN~2020faa was a more luminous event. 

In Fig. \ref{fig:sp_comp2}, we highlight the complete {\tt SYN++} models and elemental contributions calculated for +83/88 days spectra of SNe 2021aatd and 2020faa (also used for comparison in Fig. \ref{fig:sp_comp}), respectively. We also generated a {\tt SYN++} model for the +84d spectrum of SN~1987A. 
The identified ions are broadly similar in all three cases; however, modeling the blue part of the spectrum of SN~2020faa is complicated because of its lower quality. In the spectrum of SN~1987A, as mentioned above, \ion{Na}{i} D and \ion{Ba}{ii} 6142 \AA\ have significantly deeper profiles than in the cases of SNe~2021aatd and 2020faa.  Additionally, we added two further ions -- \ion{N}{ii} and \ion{O}{i} -- to the model of SN~1987A to describe its +84d spectrum.

In Fig. \ref{fig:sp_vel_comp}, beyond H$\alpha$ and \ion{Fe}{ii} 5169\AA\ line velocities, photospheric velocity ($v_{\rm phot}$) evolution of both SNe 2021aatd and 2020faa, determined from {\tt SYN++} models, is also shown.
In the case of SN~2020faa, the first $\sim$2 months are poorly covered by spectra; however, it can be clearly seen at later phases that both of its photospheric and line velocities are much higher than that of the other two objects. SN~2021aatd has, at the same time, also higher line velocities than that of SN~1987A (especially in the case of the \ion{Fe}{ii} 5169\AA\ line). 
It can be seen also that $v_{\rm phot}$ values are definitely larger than that of $v_{\text{FeII}}$ in both SNe~2021aatd and 2020faa. In SNe IIP,  $v_{\text{FeII}}$ is thought to be a good indicator of the photospheric velocity ($v_{\text{phot}}$) after $\sim$20 days, since the minimum of the \ion{Fe}{ii} 5169\AA\ absorption profile tends to form near the photosphere \citep[see][]{Branch_2003}. However, the detailed investigation of \citet{TV_12} showed that the true $v_{\text{phot}}$ may significantly differ from single line velocities. In the cases of SNe~2021aatd and 2020faa, as our {\tt SYN++} models suggest, \ion{Fe}{ii} lines can be blended e.g. with \ion{Mg}{i}.

\subsubsection{Comparison of spectral data with Dessart \& Hillier (2019) models}\label{sec:anal_sp_DH19}

\begin{figure*}
\centering
\includegraphics[width=.95\columnwidth]{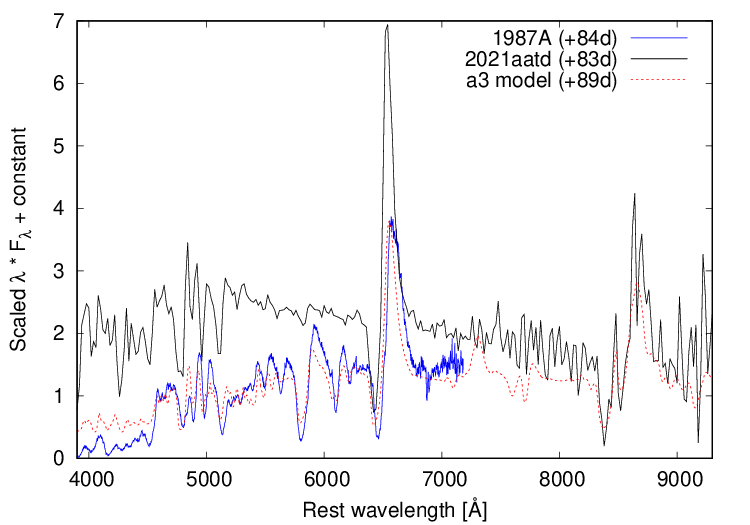}
\includegraphics[width=.95\columnwidth]{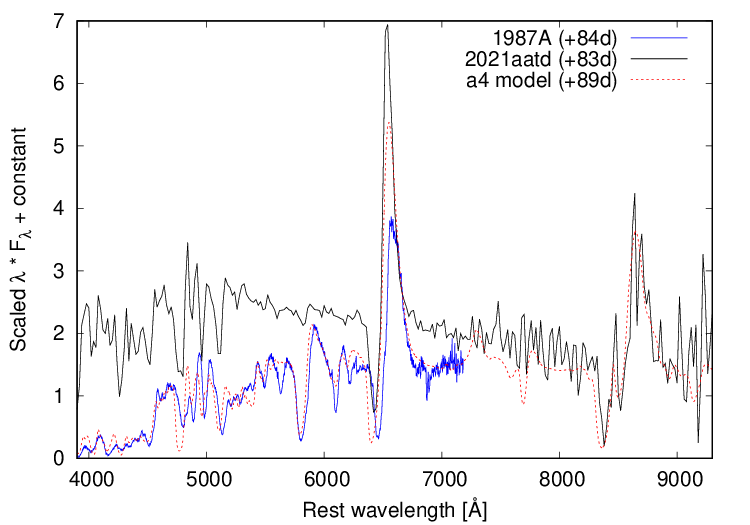}
\includegraphics[width=.95\columnwidth]{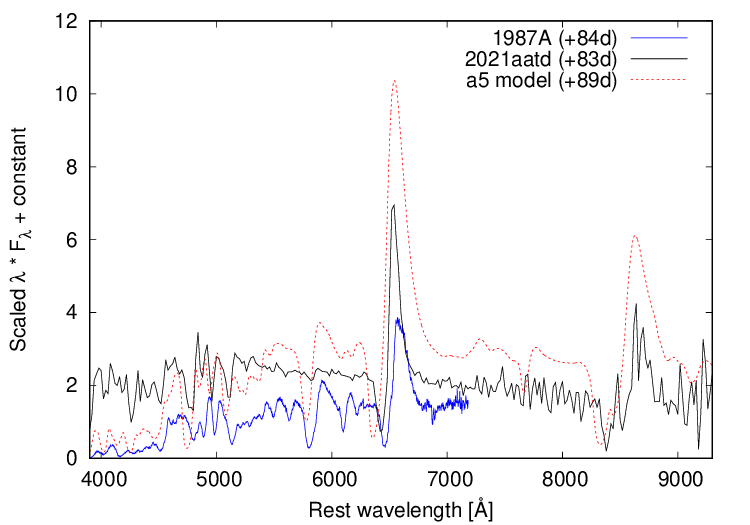}
\caption{Comparison of +83 and +84d spectra of SNe~2021aatd and 1987A to the {\it a3} (top left), {\it a4} (top right), and {\it a5} (bottom) model spectra published by \cite{Dessart_2019}; see details in text. All spectra are normalized to the flux values at 6000 \AA.}
\label{fig:sp_comp_DH19}
\end{figure*}

\begin{table}
\begin{center}
\caption{Ejecta masses, kinetic energies, and initial $^{56}$Ni masses in the adopted \citet{Dessart_2019} explosion models.}
\label{tab:DH19}
\begin{tabular}{lccc}
\hline
\hline
\smallskip
Model & $M_{\rm ej}$ & $E_{\rm kin}$ & $M_{\rm Ni}$ \\
 & ($M_{\odot}$) & (10$^{51}$ erg) & ($M_{\odot}$) \\
\hline
\smallskip
{\it a3} & 13.53 & 0.87 & 0.047 \\
{\it a4} & 13.22 & 1.26 & 0.084 \\
{\it a5} & 13.10 & 2.46 & 0.157 \\
\hline
\end{tabular}
\end{center}
\end{table}

\begin{figure*}
\centering
\includegraphics[width=.95\columnwidth]{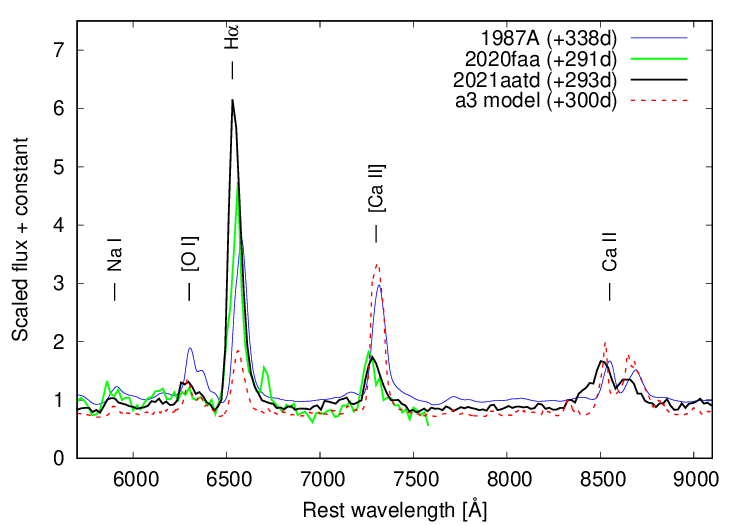}
\includegraphics[width=.95\columnwidth]{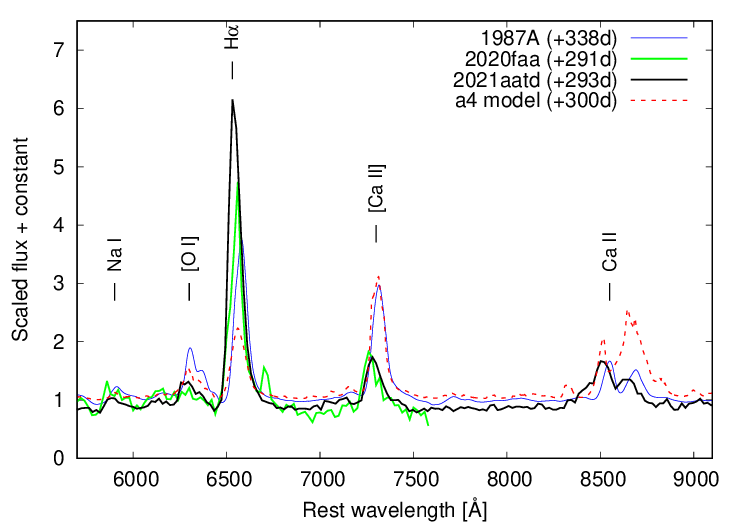}
\includegraphics[width=.95\columnwidth]{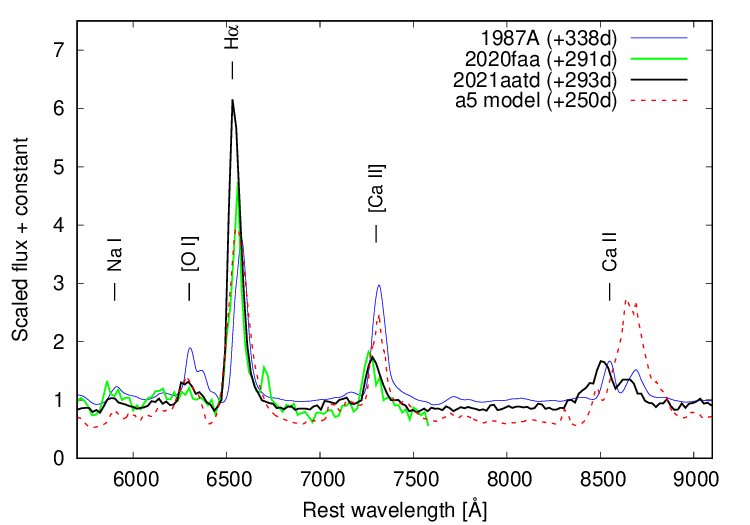}
\caption{Comparison of nebular spectra of SNe~2021aatd, 2020faa and 1987A to the {\it a3} (top left), {\it a4} (top right), and {\it a5} (bottom) model spectra published by \cite{Dessart_2019}; see details in text. All spectra are normalized to the flux values at 6000 \AA.}
\label{fig:sp_comp_neb}
\end{figure*}

We also compare our spectral dataset and the determined velocities and temperatures to the outputs of radiative transfer ({\tt cmfgen}) models published by \citet[][hereafter DH19]{Dessart_2019}. DH19 presented various sets of model spectra (covering the first $\sim$250-300 days after explosion), representing the final outcomes of the explosion of a single $M_{\rm ZAMS}$ = 15 $M_{\odot}$ low-metallicity progenitor that dies as a BSG, in order to study in detail the origin of 87A-like SNe. As the authors state, their model set was not designed to properly reproduce the observations; however, model outputs can be well used for giving constraints on certain physical parameters.

We selected three of the DH19 models ({\it a3}, {\it a4}, {\it a5}), which main parameters -- ejecta masses, kinetic energies, and initial $^{56}$Ni masses -- are shown in Table \ref{tab:DH19}. Comparison of these DH19 model spectra to the observed +83/84d spectra of SNe~2021aatd and 1987A are shown in Fig. \ref{fig:sp_comp_DH19}. As DH19 has already shown, spectra of SN~1987A can be best described with the {\it a4} model. At the same time, {\it a3} and {\it a5} models reproduce the main features and shape of the spectra; note however that {\it a5} model seems to be definitely too energetic. 

Unfortunately, available photospheric spectra of SN~1987A have a cut-off at $\sim$7200\AA\ ; nevertheless, the selected DH19 models seem to describe the red part of the spectrum of SN~2021aatd quite well. At the same time, SN~2021aatd shows a clear blue-part excess (i.e. larger temperature) that was also discussed in Sect. \ref{sec:anal_phot} regarding the comparison of both color curves and observed spectra (Figs. \ref{fig:color_comp} and \ref{fig:sp_comp}, respectively). Indeed, all DH19 models give much lower photospheric temperatures ($\sim$4300$-$4700 K) at $\sim$83 days than value of $T_{\rm phot}$=6300K we determined from the observed data using {\tt SYN++}, see Fig. \ref{fig:sp_vel_comp}. 
Model velocity curves published by DH19 (also shown in Fig. \ref{fig:sp_vel_comp}), however, strengthen the picture that both {\it a3} and {\it a4} models can be valid descriptions of the event of SN~2021aatd. Note that, at the same time, {\it a5} velocity curves are in a quite good agreement with that of SN~2020faa.

Furthermore, as DH19 have already described in detail, LCs calculated from their spectral models do not give back the proper shape of 87A-like events -- the model LCs are overall too broad and, specifically for SN1987A, {\it a4} model LC shows a slower rise to a fainter maximum. Comparing the value of the observed (second) $V$ peak magnitude of SN~2021aatd ($V_{\rm max}=-16.60 \pm  0.16$ mag, see Table \ref{tab:LC_peaks}) with the calculated $V$ peak magnitudes in DH19, there is an even larger difference: this value is much larger than that of {\it a3} and {\it a4} model LCs ($V_{\rm max} \approx -$15.0 and $-$15.6 mag, respectively), and even slightly larger than that of {\it a5} model ($V_{\rm max} \approx -$16.4 mag), see bottom right panel of Fig. \ref{fig:sp_vel_comp}. Note that the V-band LC calculated from the $^{56}$Ni-enhanced {\it a4ni} model of DH19 actually reaches the observed peak brightness of SN~2021aatd; however, its shape is even broader than that of {\it a4} and {\it a5} model LCs.

DH19 list several possible explanations for resolving the LC problem at the given 15 $M_{\odot}$ progenitor configuration -- clumping of ejecta matter, weaker $^{56}$Ni mixing, or a greater He-to-H abundance ratio --, and also note that a smaller ejecta mass would result in a faster rise to maximum and a brighter peak.
In the cases of both SNe~2021aatd and 2020faa, the remarkable early-phase flux excess needs a further explanation. 
We discuss the potential explanations of the LC evolution in Sec. \ref{sec:anal_bol}.

\subsubsection{Progenitor mass from the nebular spectrum}\label{sec:anal_sp_J14}

\begin{figure}
\centering
\includegraphics[width=.95\columnwidth]{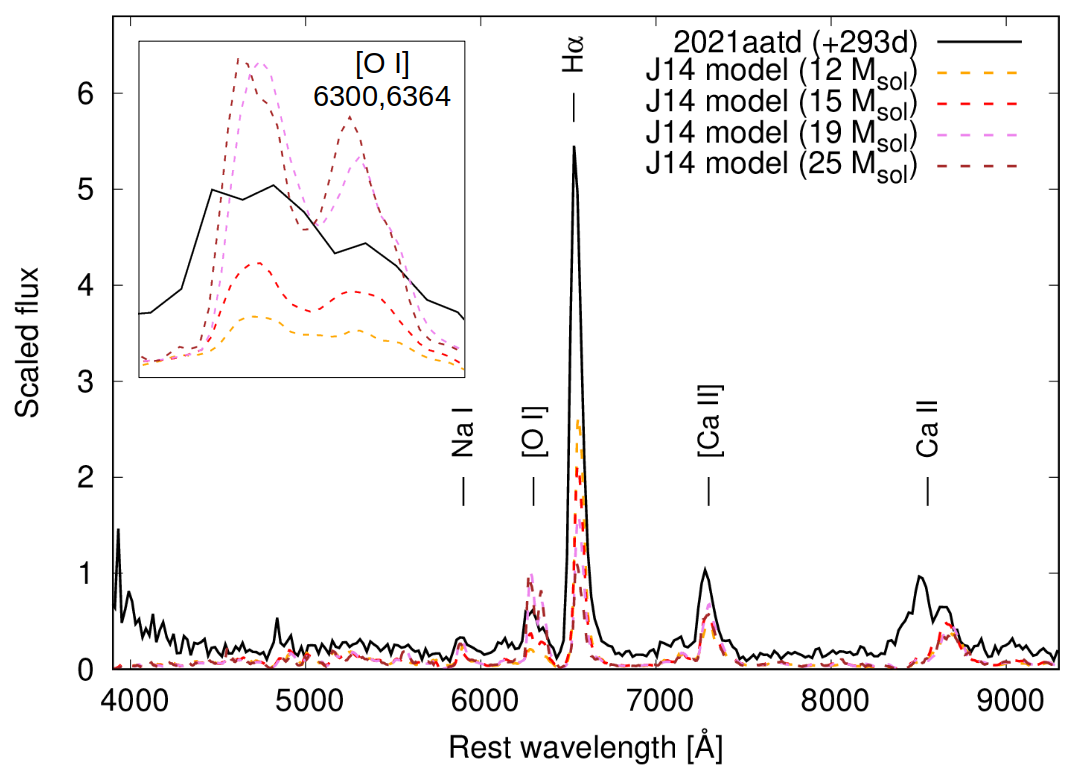}
\caption{Comparison of the +293d nebular spectrum of SN~2021aatd to the model spectra published by \cite{Jerkstrand_2012,Jerkstrand_2014}. Model spectra with the closest epochs to that of the observed one (+306 days for the 12, 15, and 25 $M_{\odot}$ models, and +332 days for the 19 $M_{\odot}$ model, respectively) have been selected; see details in text. All spectra are normalized to the flux values at 6000 \AA. The [O I] line doublet ($\lambda\lambda$6300, 6364) for the model and the observed spectra is shown in the inset; the observed line profile resembles mostly to the 15 $M_\odot$ model suggesting that the initial mass of the progenitor is close to this value.}
\label{fig:sp_comp_neb_J14}
\end{figure}

Beyond the comparison with the DH19 models, we can also use our single nebular (+293d) spectrum of SN~2021aatd for the estimation of the initial progenitor mass adopting the spectral models published by \citet{Jerkstrand_2012,Jerkstrand_2014}. 
Based on their works, a monotonic relation exists between the intensity of the [OI] doublet ($\rm{\lambda\lambda 6300, 6364}$) and the mass of the progenitor star. 
Here we compare the observed [\ion{O}{i}] emission line using the model spectra presented in \citet{Jerkstrand_2012,Jerkstrand_2014}.
The adopted synthesized spectra were produced for $\rm{M_{ZAMS}=12, 15, 19,}$ and 25 $\rm{M_{\odot}}$ at epoch of 306 days for the 12, 15, and 25 $\rm{M_{\odot}}$ models and at 332 days for the 19 $\rm{M_{\odot}}$ model, respectively. These models were originally generated for the Type IIP SN~2004et using a nickel mass of $\rm{M_{{}^{56}Ni,mod}}$ = 0.062 $\rm{M_{\odot}}$ and $\rm{d_{mod}}$ = 5.5 Mpc.
To apply these models to SN~2021aatd, the synthetic spectra are scaled to the inferred distance (62.6 Mpc) and nickel mass (0.056 $\rm{M_{\odot}}$, see Table \ref{tab:bol_lc_models}) via the following relation \citep[described e.g. in][]{szalai19b}:
\begin{equation}\label{eqn:NiMass}
F_{\rm obs} = F_{\rm mod} \times \left(\frac{d_{\rm mod}}{d_{\rm obs}}\right)^{2} \left(\frac{M_{\rm {}^{56}Ni, obs}}{M_{\rm {}^{56}Ni, mod}}\right) e^{\frac{t_{\rm mod} - t_{\rm obs}}{111.4}},
\end{equation}

\noindent where $\rm{F_{obs}}$ and $\rm{F_{mod}}$ are the observed and model fluxes, $\rm{d_{obs}}$ and $\rm{d_{mod}}$ are the observed and model distances, and $\rm{M_{{}^{56}Ni, obs}}$ and $\rm{M_{{}^{56}Ni,mod}}$ are the observed and model nickel masses synthesized during the explosion. 

The result of this comparison can be seen in Fig. \ref{fig:sp_comp_neb_J14}. We find that both the shape and the intensity level the observed line profile resemble mostly to those of the 15 $M_\odot$ model, suggesting that the initial mass of the progenitor is close to this value. This supports our previous findings (Section \ref{sec:anal_sp_DH19}) showing that spectral observables of SN~2021aatd can be well described with the explosion of a $M_{\rm ZAMS} \sim15 M_\odot$ star. 
Note that, however -- instead of cases of Type IIP SNe~2004et \citep{Jerkstrand_2012} and 2017eaw \citep{szalai19b} assumed to also arise from a $M_{\rm ZAMS} \sim15 M_\odot$ star --, the general match between model and observed spectra is not perfect. A potential reason behind that could be that Jerkstrand et al. developed these models not for studying exploding BSG but RSG stars.

\subsection{Bolometric LC modeling}\label{sec:anal_bol}

\subsubsection{Constructing the bolometric LC of SN~2021aatd}

\begin{figure}
\centering
\includegraphics[width=.95\columnwidth]{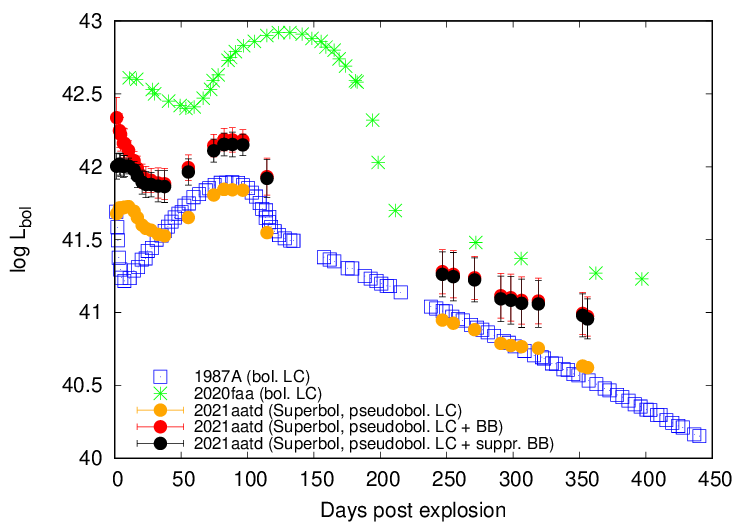}
\caption{{\it BVgri} (pseudo)bolometric LCs of SN~2021aatd calculated with the \texttt{SuperBol} code. Orange and red filled circles denote 'pure' pseudo-bolometric LC (i.e. integration of {\it BVgri} fluxes) and the one corrected after integration of a blackbody (BB)
function fitting of the available photometry, respectively. Black circles show the final bolometric LC of SN~2021aatd calculated from a suppressed BB estimation (see details in the text). Bolometric LCs of SNe~1987A and 2020faa (adopted from \citealt{nagy16} and \citealt{Salmaso_2023}, respectively) are also shown.}
\label{fig:lcbol}
\end{figure}

In principle, generating the bolometric luminosities requires sampling the spectral energy distribution along the whole electromagnetic spectrum. However, proper observational coverage in all photometric bands is not the case for SN 2021aatd.

The bolometric LC of SN 2021aatd was constructed using \texttt{SuperBol} code \citep{Nicholl_2018}. As a first step, we generated a pseudobolometric LC due to the simple integration of the observed fluxes measured in the {\it BVgri} bands. 
To do that, we used only that epochs where we have data measured in all the filters (29 out of 35 nights in total), corrected the magnitudes for extinction, and converted them to flux densities. For the integration of the flux densities, the code uses a simple trapezoidal rule and assumes zero flux below and above the limit defined by the filter equivalent width of the bluest and reddest filter, respectively.
Finally, the measured {\it BVgri} flux was converted into luminosity using the host distance given in Table \ref{tab:sne}.

Next, we made efforts on calculating the flux contribution from the unobserved ultraviolet (UV) and infrared (IR) regimes. A basic step is to fit a blackbody (BB) function to the observed flux densities to estimate the UV/IR part of the spectral energy distributions (SEDs). Generally, as presented e.g. in \citet{Salmaso_2023}, IR contribution can be well estimated with the integration of the Rayleigh-Jeans tail of the fitted BB function.

This is not the case in the UV. As also noted by \citet{Salmaso_2023} (and described in detail previously by \citealt{DH_2005}), strong depletion caused by metal lines leads depressed UV fluxes with respect to the BB function.
In the case of SN~2021aatd, no observational UV data are available to directly estimate the degree of this suppression. While, atmospheric models of 87A-like explosions calculated by \citet{Dessart_2019} do cover the UV regime, too, these probably cannot be reliably used for estimating the UV contribution in the case of SN 2021aatd: as we show in Figs. \ref{fig:color_comp} and \ref{fig:sp_comp}, early-time ($\lesssim$40 d) spectral properties and color evolution of SN~2021aatd differ quite much from those of 1987A, especially in the blue region. 
At the same time, in this earliest phase, SN~2021aatd seems to be very similar to the special Type II SN~2020faa regarding both their color curves and spectra, just at a lower luminosity scale (as we also described in detail above).

Thus, we followed a method somewhat similar to that of \citet{Salmaso_2023} in the case of SN~2020faa. Based on their Fig. 2, we adopted the UV photometry of Type II-P SN~2012aw obtained at an epoch of +28 days, shows a definite suppression regarding the BB function ($T_{BB}$=6500K) fitted to its optical-near-IR flux densities. Additionally, we also adopted the +30 days UV spectrum of SN~1987A\footnote{The spectrum was downloaded from WISeREP (https://www.wiserep.org/)}. We used the option in \texttt{SuperBol} taking into account line blanketing at UV wavelengths applying the formula $L_{UV}(\lambda < \lambda_\textrm{cutoff}) = L_{BB}(\lambda)*(\lambda/\lambda_\textrm{cutoff})^x$, where users can choose the values of the cutoff wavelength ($\lambda_\textrm{cutoff}$) and of the suppression index ($x$). We found that applying values of $\lambda_\textrm{cutoff}$=4000 \AA\ and $x$=4.0 results in a reliable shape for the suppressed BB curve comparing to the measured UV SED/spectrum of SNe~2012aw and 1987A.

Thus, the 'final' bolometric LC of SN~2021aatd was constructed using this approximation; note however, that this final luminosity curve only slightly differs from the one calculated via full BB correction after the first $\sim$2 weeks (when the largest part of the SED of the SN shifts to the optical regime). 'Final' bolometric LC of SN~2021aatd, together with the results of other approximations used in \texttt{SuperBol}, are shown in Fig. \ref{fig:lcbol}. Comparison of the final bolometric LC of SN~2021aatd to that of SN1987A and 2020faa are in agreement with the trends seen in the filtered LCs of these objects (see Fig. \ref{fig:color_comp}). 

\subsubsection{Two-component semi-analytic modeling of the bolometric LCs of SNe 1987A and 2021aatd}

\begin{figure}
\centering
\includegraphics[width=.95\columnwidth]{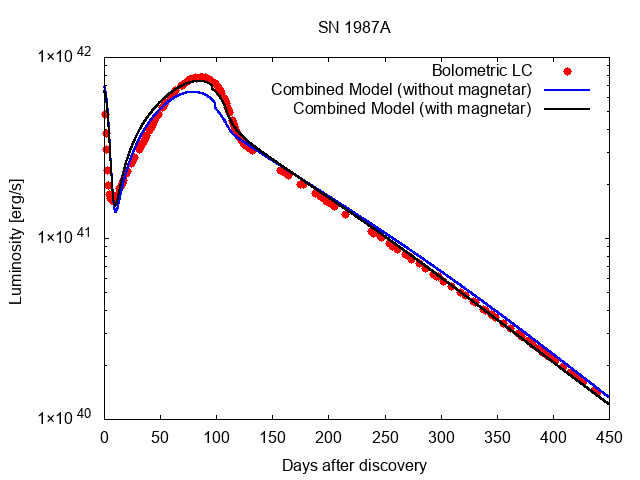}
\caption{Bolometric LC of SN~1987A (circles) with the best-fit two-component Arnett-Fu models -- assuming either only radioactive energy input (blue line, \citealt{nagy16}), or Ni-Co decay with an additional magnetar power source (black line, Model I from this work), respectively.}
\label{fig:87a_comp}
\end{figure}

\begin{figure*}
\centering
\includegraphics[width=.95\columnwidth]{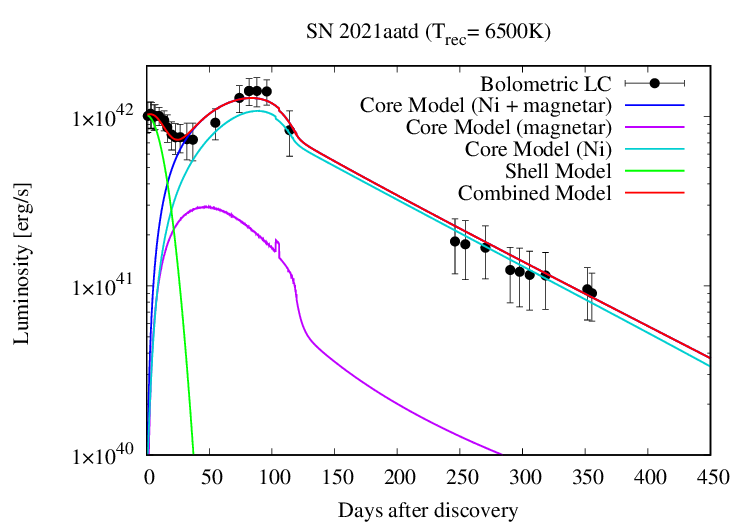}
\includegraphics[width=.95\columnwidth]{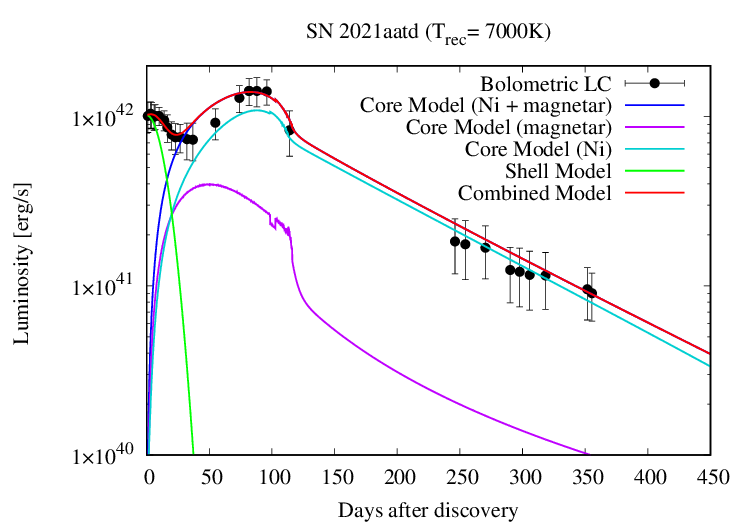}
\caption{Bolometric LC of SN 2021aatd (black circles) with the best-fit two-component Model I curves (red line); left and right panels show the results of our model fitting assuming a recombination temperature of $T_{\rm rec}$= 6500 and 7000 K, respectively. The contributions of the 'core' and the 'shell' parts of the ejecta are plotted with blue and green lines, respectively. The light blue and magenta lines represent the contribution of the radioactive decay and magnetar energy input to the luminosity evolution of the 'core' component, respectively. Details and parameter values can be found in the text and in Table \ref{tab:bol_lc_models}.}
\label{fig:lc_bolmod}
\end{figure*}


\begin{figure*}
\centering
\includegraphics[width=.95\columnwidth]{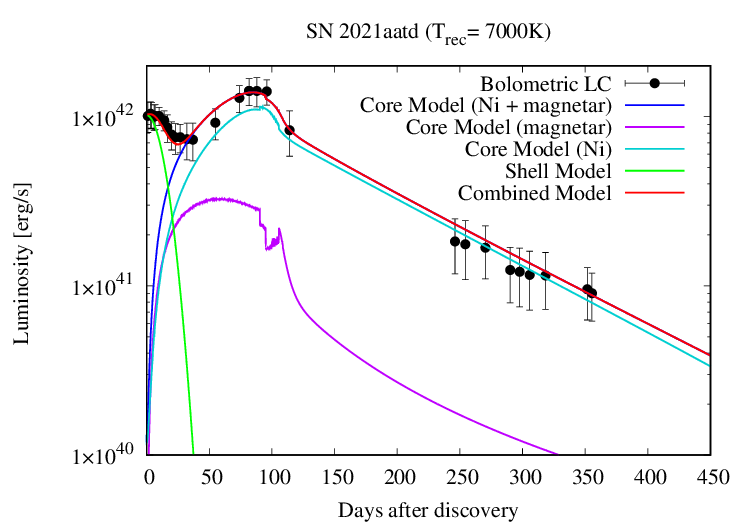}
\includegraphics[width=.95\columnwidth]{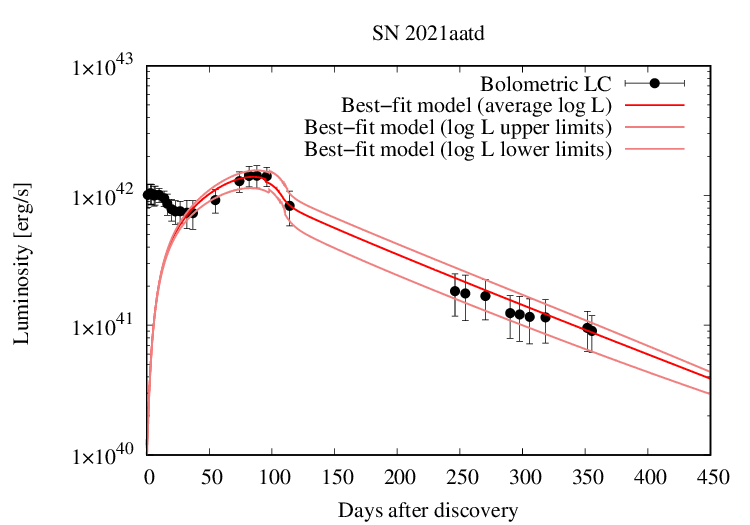}
\caption{{\bf Left:} the same as in Fig. \ref{fig:lc_bolmod}, but showing the fit of Model II to the bolometric LC of SN 2021aatd. {\bf Right:} best-fit Model II curves fitted to the average values and to the upper and lower limits of the bolometric luminosities, respectively. Details and parameter values can be found in the text and in Table \ref{tab:bol_lc_models}.}
\label{fig:lc_bolmod2}
\end{figure*}

For estimating the physical properties of SN explosions, we use a semi-analytical diffusion model originally introduced by \citet{Arnett_Fu_1989}, which assumes a homologous expanding and spherically symmetric SN ejecta with a uniform density profile and also takes into account the effect of recombination. However, the LC of SN~2021aatd shows a double peak, which cannot be fitted by the Arnett model. To deal with this cooling phase of the shock breakout, we assume a two-component structure for the SN ejecta \citep[applying a model developed by some of us, see][]{nagy14,nagy16}. Here, the first peak of the LC is dominated by the adiabatic cooling of the shock-heated, H-rich envelope (referred to as the 'shell'), and the second peak is powered by radioactive decay deposited in the denser inner region ('core'). Thus, the main difference between the two components is that the outer, low-density 'shell' is assumed to be powered only by shock heating (and not by radioactive decay).

As a starting point, we adopted the model parameters we used for modeling the bolometric LC of SN~1987A in \cite{nagy16}, hereafter NV16, assuming a moderate value of ejecta mass ($M_{\rm ej} \sim 8 M_{\odot}$) and a recombination temperature of T$_{\rm rec}$ = 5500K. Nevertheless, we note that our original model presented in NV16 is unable to fit the second ('main') LC peak of SN~1987A properly. 
For modeling the observed LCs as closely as possible, we apply here an improved version of our previously published LC fitting method: we involve an alternative energy source in addition to the radioactive decay, which considers the energy deposition of the rotational energy of a newborn magnetar \citep[just as applied by][in the case of SN~2020faa]{Salmaso_2023}. The magnetar model has the advantage of reducing the required nickel mass as well as altering the slope of the late-time LC.
Fig. \ref{fig:87a_comp} shows the comparison of the original (NV16) and the improved model without and with the magnetar contribution, respectively. It can be seen that the late decreasing part of the LC of SN~1987A can be well explained with pure Ni-Co decay, but involving the magnetar energy source i) results in a successful modeling of the main LC peak at roughly +90 days, ii) does not affect the early-time LC fit ('shell' component), and iii) allows the fine-tuning of the fit of the late-time LC. 

Thus, we decided to continue the LC modeling process applying this improved method (called hereafter Model I). The main model parameters -- beyond recombination temperature (T$_{\rm rec}$) and ejecta mass ($M_{\rm ej}$) mentioned above -- are: the progenitor radius $R_0$ (in 10$^{11}$ cm units), the initial mass of the radioactive $^{56}$Ni ($M_{\rm Ni}$, in $M_\odot$), the total energy ($E_{\rm tot}$, in 10$^{51}$ erg), the ratio of the kinetic and thermal energies of the ejecta ($E_{\rm kin}/E_{\rm Th}$), the average opacity of the ejecta ($\kappa$, in cm$^2$/g), and the spin-down energy of the magnetar ($E_{\rm p}$, in 10$^{51}$ erg) and its characteristic time-scale ($t_{\rm p}$, in days).
As can be seen in Table \ref{tab:bol_lc_models}, 'Model I' LC parameter set of SN~1987A show some differences regarding that of previously published in NV16. The most significant changes are in the values of T$_{\rm rec}$ (7000 vs. 5500 K), and in 'core' parameters $R_{\rm 0,core}$ (2.8 vs. 29.0 $\times 10^{11}$ cm) and $E_{\rm kin, core}/E_{\rm Th, core}$ (3.2 vs. 11.7). 
At the same time, it was not necessary to change any 'shell' parameters.

As a next step, we fit Model I to the final bolometric LC of SN~2021aatd calculated with \texttt{SuperBol} (see above). In Fig. \ref{fig:lc_bolmod}, we present two slightly different solutions assuming recombination temperatures of $T_{\rm rec}$ = 6500 K and 7000 K (left and right panels, respectively).
The reason of that is the presence of a modest re-brightening on the bolometric LC around +300 days, which may be a hint for CSM interaction -- but the late part of the LC is not well-covered enough to consider it as a strong evidence. Nevertheless, late-time LC evolution of SN~2021aatd is not linear and cannot be perfectly described with one model. Thus, we decided to take into account both cases: the left panel of Fig. \ref{fig:lc_bolmod} ($T_{\rm rec}$ = 6500~K) shows the best fit for data up to $\sim$300 days, while the right panel ($T_{\rm rec}$ = 7000~K) is the best solution for fitting including the latest data points, too. 
Note, however, that both parameter sets represent a hydrogen-depleted ejecta and that they are quite similar; there are only small differences in some 'core' parameters but no changes in 'shell' parameters at all (see Table \ref{tab:bol_lc_models}). In both panels of Fig. \ref{fig:lc_bolmod}, we also present the contributions of both the radioactive decay and magnetar energy input to the 'core' luminosity component, respectively.

\begin{table*}
\caption{Parameters of the best-fit two-component bolometric LC models for SNe 2021aatd and 1987A. See text for a detailed description.}
\label{tab:bol_lc_models}
\centering
\renewcommand{\arraystretch}{2}
\begin{tabular}{l||cc|c||c|c|c} 
\hline 
\hline
 & \multicolumn{3}{c||}{SN 2021aatd}  & \multicolumn{3}{c}{SN 1987A} \\ \hline
 & \multicolumn{2}{c|}{Model I} & Model II & NV16 & Model I & Model II \\
T$_{\rm rec}$ (K) & 6500 K & 7000 K & 7000 $ \pm$ 500 K & 5500 K & 7000K & 7000 K\\ \hline
& \multicolumn{6}{c} {'Core'} \\ \hline
$R_0$ (10$^{11}$ cm) & 1.8 & 1.3 & 0.64 $\substack{+0.1 \\ -0.2}$ & 29.0 & 2.8 & 1.3\\
$M_{\rm ej}$ ($M_\odot$) & 7.4 & 7.7 & 14.35 $\substack{+0.88 \\ -1.12}$ & 8.6 & 7.9 & 14.35\\
$M_{\rm Ni}$ ($M_\odot$) & 0.13 & 0.13 & 0.13 $\substack{+0.03 \\ -0.05}$ & 0.069 & 0.058 & 0.058\\
$E_{\rm tot}$ (10$^{51}$ erg) & 1.75 & 1.7 & 6.4 $\substack{+0.7 \\ -1.5}$ & 1.5 & 1.7 & 3.61\\
$E_{\rm kin}$/E$_{\rm Th}$  & 2.5 & 3.25 & 1.2 $\substack{+0.03 \\ -0.11}$ & 11.7 & 3.2 & 1.91 \\
$\kappa$ (cm$^2$/g) & 0.3 & 0.3 & 0.24 & 0.19 & 0.20 & 0.21 \\  
$E_{\rm p}$ (10$^{51}$ erg) & 0.012 & 0.013 & 0.016 $\substack{+0.006 \\ -0.002}$ & -- & 0.014 & 0.014\\
$t_{\rm p}$ (days) & 6.0 & 7.0 & 7.0 $\pm$ 1.0 & -- & 6.0 & 6.0\\ \hline
 & \multicolumn{6}{c} {'Shell'} \\ \hline
$R_0$ (10$^{13}$ cm) & \multicolumn{3}{c||}{1.5 $\pm$ 0.3}& \multicolumn{2}{c|}{1.0} & 1.5\\
$M_{\rm ej}$ ($M_\odot$) & \multicolumn{3}{c||}{0.3} & \multicolumn{2}{c|}{0.1} & 0.05 \\
$M_{\rm Ni}$ ($M_\odot$) & \multicolumn{3}{c||}{ -- } & \multicolumn{2}{c|}{--} & --\\
$E_{\rm tot}$ (10$^{51}$ erg) & \multicolumn{3}{c||}{0.303 $\substack{+0.002 \\ -0.013}$} & \multicolumn{2}{c|}{0.4} & 0.095 \\
$E_{\rm kin}$/E$_{\rm Th}$ & \multicolumn{3}{c||}{2.26 $\substack{+0.01 \\ -0.05}$} & \multicolumn{2}{c|}{20.0} & 18.0\\
$\kappa$ (cm$^2$/g) &\multicolumn{3}{c||}{0.34} & \multicolumn{2}{c|}{0.34}& 0.34\\  
\hline
\end{tabular}
\tablefoot{Parameters are: the recombination temperature (T$_{\rm rec}$), the progenitor radius $R_0$, the ejecta mass ($M_{\rm ej}$), the initial mass of the radioactive $^{56}$Ni ($M_{\rm Ni}$), the total energy ($E_{\rm tot}$), the ratio of the kinetic and thermal energies of the ejecta ($E_{\rm kin}/E_{\rm Th}$), the average opacity of the ejecta ($\kappa$), and the spin-down energy of the magnetar ($E_{\rm p}$) and its characteristic time-scale ($t_{\rm p}$).}
\end{table*}

Best-fit parameters sets of 'Model I' for the data of both SNe 1987A and 2021aatd give an ejecta mass of $M_{\rm ej}\sim 8 M_\odot$, similarly to the original conclusion of NV16. However, taking into account both our findings on comparing observed and model spectra (Sections \ref{sec:anal_sp_DH19} and \ref{sec:anal_sp_J14}) and the results of previous works on stellar evolution and explosion processes of BSG stars \citep[e.g.][]{Blinnikov_2000,Utrobin_2015,Dessart_2019,Xiang_2023}, it seems to be plausible to assume a larger ejecta mass. This problem was already discussed in NV16; from the point of view of the applied LC modeling method, the key is the degeneration of ejecta mass ($M_{\rm ej}$) and average opacity ($\kappa$). As described in NV16 and references therein in detail, for H-rich CCSNe, the potential range of the average opacity in the core is $\kappa_{\rm core} \approx 0.2 \pm 0.1$ cm$^2$/g; if one let $\kappa_{\rm core}$ vary within the whole interval, value of $M_{\rm ej, core}$ (and, thus, mass of the progenitor) may constrained within a factor of $\sim$2. There is the same relation between $\kappa_{\rm shell} \approx 0.3-0.4$ and $M_{\rm ej, shell}$; however, since $M_{\rm ej, shell}$ has a much lower value, it contributes less to the mass of the progenitor star. 

%
Thus, as a next step, we defined 'Model II', fixing $T_{\rm rec}$ = 7000 K and $M_{\rm ej, core}$ = 14.35 $M_\odot$ for both SNe~2021aatd and 1987A; these, together with $M_{\rm ej, shell} \approx 0.1-0.2 M_\odot$, are in a very good agreement with the assumptions of the atmospheric models {\it a3} and {\it a4} of \citet{Dessart_2019}. As 'Model II' seems to be the more plausible solution, we estimated the uncertainty of the model parameters for SN 2021aatd. To do so, we create models and fit them to not only the average values but also to the upper and lower limits of the bolometric luminosities. At the same time, it was necessary to change the 'shell' parameters for both SNe 2021aatd and 1987A to get similarly good fits as in the case of 'Model I'; see Fig. \ref{fig:lc_bolmod2} and Table \ref{tab:bol_lc_models} for the results. During the fitting of 'Model II', we used a power-law density profile (using an index of n=1.8) for SN 2021aatd, while we got a better fit by assuming a constant density profile for SN 1987A.
While core Ni-mass and magnetar parameters remain unchanged, we get smaller values for the radius and larger values for the total energy of the core component; $\kappa_{\rm core}$, being 0.24 and 0.21 cm$^2$/g for SN~2021aatd and 1987A, respectively, is still in the desired range. 

\begin{figure}
\centering
\includegraphics[width=.95\columnwidth]{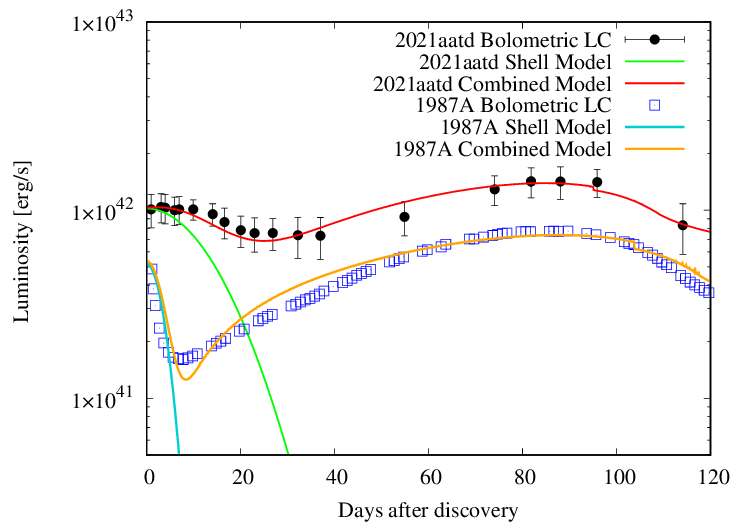}
\caption{Comparison of the early-phase 'Model II' bolometric LC fits of SNe 2021aatd and 1987A.}
\label{fig:lcbol_early}
\end{figure}

Overall, 'core' parameters of SNe 2021aatd and 1987A seem to be similar using either 'Model I' or 'Model II' configurations -- this is exactly what one expects from the similar late-time LC evolution of the two objects. However, SN~2021aatd was a more luminous event than SN~1987A, which is reflected in the values of Ni-mass (0.13 vs 0.06 $M_{\odot}$) and, in 'Model II', also of the total energy (6.4 vs. 3.6 $\times 10^{51}$~erg). At the same time, the 'core' radius of SN~2021aatd seems to be significantly smaller than that of SN~1987A.

Taking a look at the final 'shell' parameters, differences between the two events are more remarkable, just as suggested by the differences in their early-time luminosity and spectral properties (shapes of the shell LC components of both SNe can be seen better in Fig. \ref{fig:lcbol_early}). This is especially true, if we compare the 'Model II' 'shell' parameters of the objects -- based on that parameter sets, we definitely see a more massive and more energetic shell in the case of SN~2021aatd. 
A further evidence on the reliability of 'Model II' is that we get back a larger kinetic energy (and, thus, larger ejecta velocities) in the shell for SN~2021aatd than for SN~1987A, in agreement with the results we received from the analysis of the spectral datasets (see Sec. \ref{sec:anal_sp}) -- while this is not the case if we apply 'Model I'.


\section{Discussion \& Conclusions}\label{sec:concl}

Based on our detailed spectral and LC analysis, the LC evolution of SN~2021aatd after the first $\sim$40 days is very similar to that of SN~1987A. We found that this phase can be well described with the explosion of a $M_{\rm ZAMS} \sim$15 $M_\odot$ BSG star, as it was previously assumed and investigated in detail for 87A-like explosions by \citet{Dessart_2019}. However, as we noted above, photospheric temperatures are considerably higher for SN~2021aatd than for {SN~}1987A. Moreover, \ion{Na}{i} D and \ion{Ba}{ii} 6142 \AA\ lines, which appear as dominant lines in the spectra of SN~1987A, are very weak in the case of SN~2021aatd. These properties of SN~2021aatd, just as the relatively high $v_{\rm H\alpha}$ and $v_{\rm FeII}$ velocity curves shown in Fig. \ref{fig:sp_vel_comp}, resemble rather to SN IIP-like atmospheres. Nevertheless, such transitions between 1987A- and IIP-like SNe have been already described in several works \citep[e.g.][]{Takats_2016,Dessart_2019,Xiang_2023}.  

As described in Sect. \ref{sec:anal_sp_DH19}, the proper explanation of the LC evolution of 87A-like SNe based on the outcomes of radiative transfer spectral models is challenging. Extended efforts have been made for modeling bolometric LC of 1987A and of similar events using hydrodynamic codes, taking into account various aspects like rotation, mixing of $^{56}$Ni, clumping of ejecta matter, or the presence of a binary companion (see a review in \citealt{Dessart_2019}, and even more recent results in e.g. \citealt{Dessart_2019b}, \citealt{Menon_2019}, or \citealt{Utrobin_2021}). 

In this work (see Sect. \ref{sec:anal_bol}), we present a more simple, semi-analytical model assuming a two-component ejecta -- developed by some of us \citep{nagy14,nagy16}, based on the method originally introduced by \citet{Arnett_Fu_1989} --, which seems to be also appropriate for describing the bolometric LC of 87A-like SNe. While this kind of LC modeling of SN~1987A has been already published in \citet{nagy16}, for achieving an even more proper fit, we apply here an improved version of the LC model: involving the rotational energy of a newborn magnetar in addition to radioactive decay allows us to fine-tune the fitting at both around the peak and at later phases.

In the case of the primary object of this study, SN~2021aatd, we can also successfully apply this improved bolometric LC model. While, as mentioned, its late-time LC evolution is very similar to that of SN~1987A (which leads to very similar model parameters regarding the 'core' ejecta component), the early-time peculiar LC of SN~2021aatd can be also well described with changing the outer ('shell') parameters towards a larger ejecta mass and total energy than in the case of SN~1987A, choosing our 'Model II' configuration (which has a fixed recombination temperature of $T_{\rm rec}$ = 7000 K and a fixed core ejecta mass of $M_{\rm ej, core}$ = 14.35 $M_\odot$ for both SNe~2021aatd and 1987A).
Using this parameter set, we find that summed ('core' + 'shell') values of key parameters ($M_{\rm ej} \approx 14.6 M_\odot$, $M_{\rm Ni} \approx$0.13 $M_\odot$, and $E_{\rm kin} = 3.7 \times$ 10$^{51}$ erg) harmonize well with those used in the adopted DH19 spectral models for the case of SN~2021aatd (see Table \ref{tab:DH19}). 

We also note that the parameter set belongs to 'Model I' -- assuming a lower total ejecta mass of $\sim 8 M_\odot$, but a somewhat larger core radius but smaller energies, without any change in 'shell' parameters -- also gives a proper LC solution for SN~2021aatd (and also for SN~1987A, but with modified 'shell' parameters, see Table \ref{tab:bol_lc_models}). This is probably due to the known degeneration of our model between the mass and the average opacity of the ejecta \citep[see][]{nagy16}. 
Furthermore, as noted by \cite{Dessart_2019}, assuming a low ejecta mass (and, thus, a progenitor mass of $<$15 $M_\odot$) could solve the LC fitting problems of their radiative transfer models.
However, we note that the 'shell' parameters resulted by fitting 'Model I' do not reflect the true energy/velocity relations between SNe 2021aatd and 1987A (while 'Model II' parameters do).
Moreover, providing another argument in favor of a $\sim$15 $M_\odot$ progenitor, we refer here to the results of the comparison of the nebular spectrum of SN~2021aatd with the models of \citet{Jerkstrand_2012,Jerkstrand_2014} (see Sect. \ref{sec:anal_sp_J14}).

We also showed that both spectral and early-time LC evolution of SN~2021aatd (i.e. before $\sim$90 days) resembles remarkably that of the long-lived SN~2020faa (see Sect. \ref{sec:anal_phot}). Taking a look again at Figs. \ref{fig:lc_comp}$-$\ref{fig:color_comp}, the striking similarity of light and color curves of these two events during the first $\sim$3 months calls for a similar physical background; even that the luminosity scales of the two SNe differ by a factor of $\sim$5-10.
For describing the pseudo-bolometric ({\it g'+r'+i'}) LC of SN~2020faa, \cite{Salmaso_2023} also apply the \cite{nagy16} model with the built-in magnetar energy component, just as we do in the case of SNe~2021aatd and 1987A. They successfully model the early LC of SN~2020faa, however, assuming a more massive ($M_{\rm ej} \approx 23 M_\odot$) and larger ($R_{\rm 0, shell} \approx 10^{14}$ cm) progenitor.
At the same time, this model does not provide a satisfactory power explanation for the peak without an unphysically large value of Ni-mass, or, assuming magnetar spin-down parameters that are in contrast with the rapid drop of LC after the peak (which, as can be seen in Figs. \ref{fig:lc_comp} and \ref{fig:color_comp}, is quite different from the post-peak evolution of both SNe~2021aatd and 1987A). Thus, in order to describe the peak luminosity, \cite{Salmaso_2023} involved further potential explanations (hidden interaction with an inner disc, or, effects of a delayed, choked jet); however, proving these alternatives requires further investigations.

Nevertheless, our analysis on 2021aatd, as well as on SNe 1987A and 2020faa, seems to strengthen that we do see similar kind of explosions, which all can be described with a two-component ejecta (assuming extra luminosity source(s) in the case of SN~2020faa). Up to now, only a handful of 1987A-like SNe and even less long-lived, multipeaked SNe are known; thus, their connection is not clear now.
However, ongoing and near-future large-scale transient surveys are expected to find more and more similar objects to that of SN~2021aatd, helping to give a clearer view of the reason behind the similarities and diversity of exploding H-rich massive stars. 


\begin{acknowledgements}
We used WISeREP (https://www.wiserep.org/) for downloading previously published spectra. We gratefully thank Irene Salmaso for sharing the nebular spectra of SN~2020faa with us.
This work makes use of the Las Cumbres Observatory global telescope network.  The LCO group is supported by NSF grants AST-1911151 and AST-1911225.
This project has been supported by the GINOP-2-3-2-15-2016-00033 project of the National Research, Development and Innovation (NRDI) Office of Hungary funded by the European Union, as well as by NKFIH OTKA FK-134432, PD-134434 and K-142534 grants, and from the HUN-REN Hungarian Research Network. 
T.S. is supported by the J{\'a}nos Bolyai Research Scholarship of the Hungarian Academy of Sciences. R.K.T., T.S., and D.B. are supported by the \'UNKP 22-4 and \'UNKP 22-5 New National Excellence Programs of the Ministry for Culture and Innovation from the source of the NRDI Fund of Hungary, respectively.
\end{acknowledgements}

%
%
\bibliographystyle{aa}
\bibliography{main} 

\clearpage

\begin{appendix} 

\onecolumn
\section{Observational data}

\begin{center}
\begin{longtable}{cccccc}
\caption{$BVgri$ photometry of SN~2021aatd obtained from LCO sites} \label{tab:phot_data_LCO} \\
\hline
\hline
\multicolumn{1}{c}{MJD} & \multicolumn{1}{c}{B (mag)} & \multicolumn{1}{c}{V (mag)} & \multicolumn{1}{c}{g (mag)} & \multicolumn{1}{c}{r (mag)} & \multicolumn{1}{c}{i (mag)} \\ \hline 
59496.0 & 17.93 (0.07) & 17.97 (0.06) & 17.81 (0.03) & 18.03 (0.03) & 18.26 (0.04) \\
59498.1 & 17.93 (0.06) & 17.86 (0.03) & 17.73 (0.03) & 17.87 (0.03) & 18.07 (0.03) \\
59499.0 & 17.95 (0.08) & 17.88 (0.04) & 17.75 (0.04) & 17.84 (0.02) & 18.05 (0.03) \\
59501.0 & 18.02 (0.07) & 17.89 (0.03) & 17.78 (0.04) & 17.78 (0.02) & 17.98 (0.03) \\
59501.9 & 18.01 (0.06) & 17.88 (0.03) & 17.78 (0.04) & 17.76 (0.02) & 17.95 (0.02) \\
59504.9 & 18.09 (0.06) & 17.83 (0.03) & 17.86 (0.04) & 17.69 (0.03) & 17.86 (0.03) \\
59509.1 & 18.35 (0.09) & 17.91 (0.05) & 17.98 (0.06) & 17.70 (0.04) & 17.80 (0.05) \\
59511.6 & 18.56 (0.08) & 18.03 (0.03) & 18.08 (0.03) & 17.77 (0.04) & 17.92 (0.03) \\
59515.1 & 18.77 (0.07) & 18.12 (0.03) & 18.27 (0.03) & 17.87 (0.02) & 18.00 (0.03) \\
59518.0 & 18.91 (0.06) & 18.16 (0.03) & 18.37 (0.03) & 17.90 (0.03) & 18.03 (0.04) \\
59521.9 & 18.98 (0.06) & 18.21 (0.04) & 18.43 (0.04) & 17.92 (0.03) & 17.98 (0.03) \\
59527.2 & 19.20 (0.09) & 18.24 (0.03) & 18.56 (0.03) & 17.96 (0.03) & 18.03 (0.04) \\
59532.0 & 19.26 (0.09) & 18.24 (0.03) & 18.58 (0.04) & 18.02 (0.03) & 18.02 (0.04) \\
59549.9 & 18.77 (0.08) & 17.94 (0.04) & 18.17 (0.03) & 17.76 (0.03) & 17.77 (0.04) \\
59562.8 & 18.43 (0.11) & 17.73 (0.08) & 17.91 (0.08) & 17.49 (0.07) & ... \\
59569.1 & 18.32 (0.06) & 17.57 (0.03) & 17.79 (0.04) & 17.37 (0.03) & 17.42 (0.03) \\
59576.8 & 18.23 (0.07) & 17.47 (0.03) & 17.68 (0.03) & 17.28 (0.03) & 17.30 (0.05) \\
59583.1 & 18.28 (0.11) & 17.50 (0.03) & 17.68 (0.03) & 17.28 (0.03) & 17.29 (0.04) \\
59590.9 & 18.25 (0.16) & 17.47 (0.07) & 17.70 (0.07) & 17.32 (0.09) & 17.29 (0.06) \\
59605.1 & 19.20 (0.15) & 17.93 (0.08) & 18.34 (0.06) & ... & 17.72 (0.09) \\
59609.1 & 19.39 (0.10) & 18.26 (0.05) & 18.63 (0.05) & 17.86 (0.04) & 17.89 (0.05) \\
59731.2 & 20.75 (0.10) & 19.92 (0.06) & 20.13 (0.05) & ... & 19.57 (0.09) \\
59741.1 & 20.78 (0.13) & 19.91 (0.07) & 20.21 (0.06) & 19.18 (0.04) & 19.61 (0.07) \\
59749.4 & 20.86 (0.14) & 20.07 (0.11) & 20.32 (0.09) & 19.18 (0.06) & 19.63 (0.09) \\
59765.3 & 20.99 (0.11) & 20.12 (0.05) & 20.48 (0.04) & 19.34 (0.03) & 19.65 (0.05) \\
59785.2 & 21.02 (0.10) & 20.43 (0.07) & 20.67 (0.06) & 19.56 (0.04) & 19.97 (0.15) \\
59792.7 & 21.15 (0.09) & 20.47 (0.14) & 20.76 (0.06) & 19.56 (0.03) & 20.02 (0.06) \\
59800.7 & 21.24 (0.16) & 20.37 (0.07) & 20.71 (0.08) & 19.59 (0.04) & 20.15 (0.07) \\
59813.3 & 21.15 (0.08) & 20.46 (0.06) & 20.82 (0.06) & 19.61 (0.03) & 20.10 (0.05) \\
59846.7 & 21.47 (0.09) & 20.85 (0.06) & 21.00 (0.05) & 20.02 (0.03) & 20.16 (0.06) \\
59850.4 & 21.49 (0.10) & 20.82 (0.06) & 20.99 (0.05) & 20.06 (0.04) & 20.26 (0.06) \\
59865.3 & 21.57 (0.21) & ... & 21.13 (0.15) & 20.21 (0.08) & 20.50 (0.16) \\
59895.2 & 21.68 (0.20) & ... & ... & 20.44 (0.08) & 20.68 (0.31) \\
\hline
\end{longtable}
\end{center}

\newpage
\begin{center}
\begin{longtable}{cccccc}
\caption{$BVgri$ photometry of SN~2021aatd obtained from Baja Observatory} \label{tab:phot_data_Baja} \\
\hline
\hline
\multicolumn{1}{c}{MJD} & \multicolumn{1}{c}{B (mag)} & \multicolumn{1}{c}{V (mag)} & \multicolumn{1}{c}{g (mag)} & \multicolumn{1}{c}{r (mag)} & \multicolumn{1}{c}{i (mag)} \\ \hline 
59501.9 & ... & 17.94 (0.04) & 17.83 (0.09) & 17.78 (0.03) & 17.91 (0.04) \\
59510.9 & 18.59 (0.23) & 18.01 (0.06) & ... & 17.77 (0.03) & 17.78 (0.14) \\
59511.9 & 18.57 (0.24) & 18.00 (0.08) & 18.17 (0.07) & 17.78 (0.04) & 17.85 (0.05) \\
59512.8 & 18.69 (0.21) & 18.05 (0.05) & 18.17 (0.07) & 17.77 (0.02) & ... \\
59514.9 & 18.78 (0.21) & 17.99 (0.05) & 18.20 (0.04) & 17.87 (0.02) & 17.93 (0.04) \\
59515.9 & 18.73 (0.18) & 18.14 (0.06) & 18.30 (0.04) & 17.96 (0.02) & 18.03 (0.03) \\
59516.9 & 18.89 (0.21) & 18.09 (0.08) & 18.36 (0.04) & 17.92 (0.02) & 18.00 (0.04) \\
59517.9 & 18.95 (0.21) & 18.21 (0.06) & 18.34 (0.05) & 17.97 (0.03) & 17.96 (0.03) \\
59518.9 & 19.03 (0.18) & 18.22 (0.06) & 18.33 (0.06) & 17.92 (0.02) & 18.08 (0.03) \\
59538.8 & ... & 18.18 (0.09) & 18.47 (0.05) & 17.94 (0.03) & 17.85 (0.03) \\
59539.8 & ... & 18.22 (0.05) & 18.48 (0.09) & 17.97 (0.05) & 17.89 (0.05) \\
59555.8 & 18.62 (0.23) & 17.85 (0.03) & 18.04 (0.05) & 17.61 (0.02) & 17.68 (0.04) \\
59565.8 & ... & ... & 17.82 (0.24) & 17.39 (0.13) & 17.46 (0.07) \\
59585.8 & 18.28 (0.19) & 17.48 (0.06) & 17.73 (0.05) & 17.32 (0.02) & 17.29 (0.02) \\
59586.7 & 18.25 (0.21) & 17.47 (0.03) & 17.69 (0.04) & 17.29 (0.02) & 17.24 (0.04) \\
59591.7 & 18.26 (0.26) & 17.66 (0.08) & 17.73 (0.07) & 17.31 (0.05) & 17.30 (0.04) \\
59593.7 & 18.38 (0.23) & 17.60 (0.10) & 17.75 (0.08) & 17.37 (0.05) & 17.32 (0.05) \\
59594.7 & 18.37 (0.21) & 17.67 (0.07) & 17.88 (0.05) & 17.42 (0.03) & 17.40 (0.05) \\
59597.7 & ... & 17.71 (0.16) & 17.99 (0.05) & 17.45 (0.05) & 17.48 (0.05) \\
59598.8 & ... & 17.77 (0.06) & 18.09 (0.07) & 17.43 (0.06) & 17.45 (0.03) \\
59603.7 & 19.00 (0.17) & 17.92 (0.07) & 18.21 (0.04) & 17.67 (0.07) & 17.62 (0.04) \\
\hline
\end{longtable}
\end{center}

\centering
\begin{longtable}{ccccc}
\caption{Log of spectroscopic observations}
\label{tab:spec} \\
\hline
\hline
\multicolumn{1}{c}{UT Date} & \multicolumn{1}{c}{Phase} & \multicolumn{1}{c}{Instrument} & \multicolumn{1}{c}{Range} & \multicolumn{1}{c}{R} \\
~ & (days) & ~ & (\AA) & ($\lambda$/$\Delta\lambda$) \\
\hline
2021-10-09 & +2 & LCO FLOYDS & 3250$-$10\,000 & 400-700 \\
2021-10-11 & +4 & LCO FLOYDS & 3250$-$10\,000 & 400-700 \\
2021-10-13 & +6 & LCO FLOYDS & 3250$-$10\,000 & 400-700 \\
2021-10-21 & +14 & LCO FLOYDS & 3250$-$10\,000 & 400-700 \\
2021-10-26 & +19 & LCO FLOYDS & 3250$-$10\,000 & 400-700 \\
2021-10-31 & +24 & LCO FLOYDS & 3250$-$10\,000 & 400-700 \\
2021-11-09 & +33 & LCO FLOYDS & 3250$-$10\,000 & 400-700 \\
2021-11-20 & +44 & LCO FLOYDS & 3250$-$10\,000 & 400-700 \\
2021-12-16 & +70 & LCO FLOYDS & 3250$-$10\,000 & 400-700 \\
2021-12-29 & +83 & LCO FLOYDS & 3250$-$10\,000 & 400-700 \\
2022-01-13 & +98 & LCO FLOYDS & 3250$-$10\,000 & 400-700 \\
2022-07-26 & +293 & Magellan LDSS-3 & 3700$-$10\,650 & ? \\
\hline
\end{longtable}

\section{Results of SYN++ modeling}
\subsection{SN~2021aatd}

\begin{figure*}
\centering
\includegraphics[width=5.5cm]{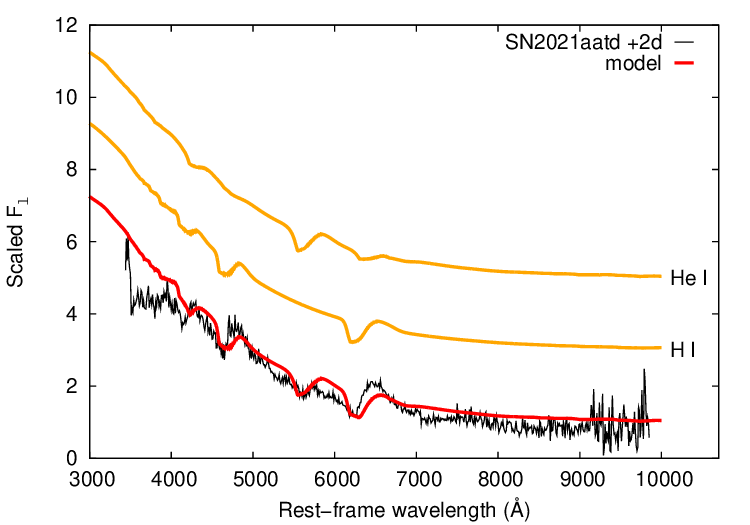}
\includegraphics[width=5.5cm]{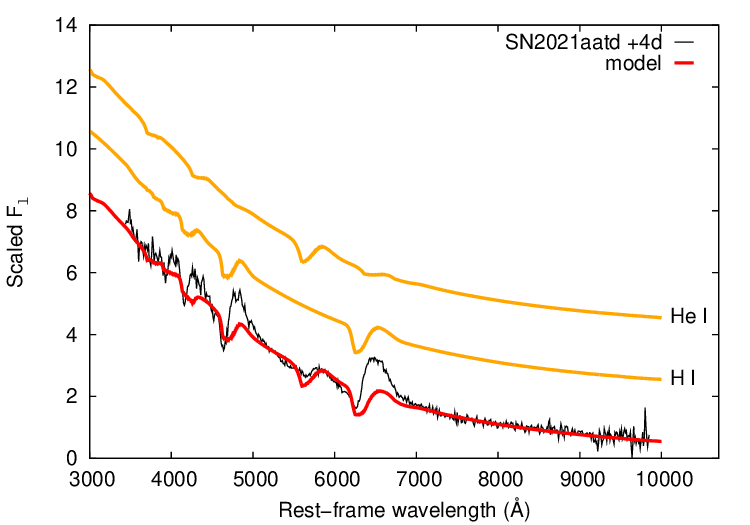}
\includegraphics[width=5.5cm]{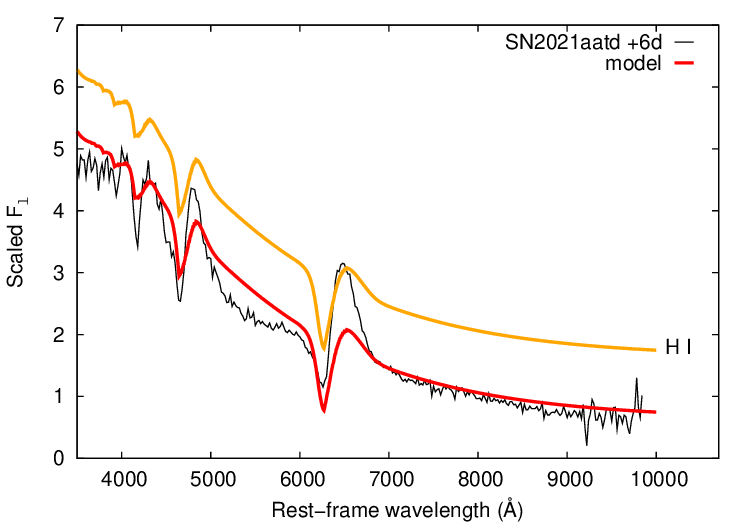}
\includegraphics[width=5.5cm]{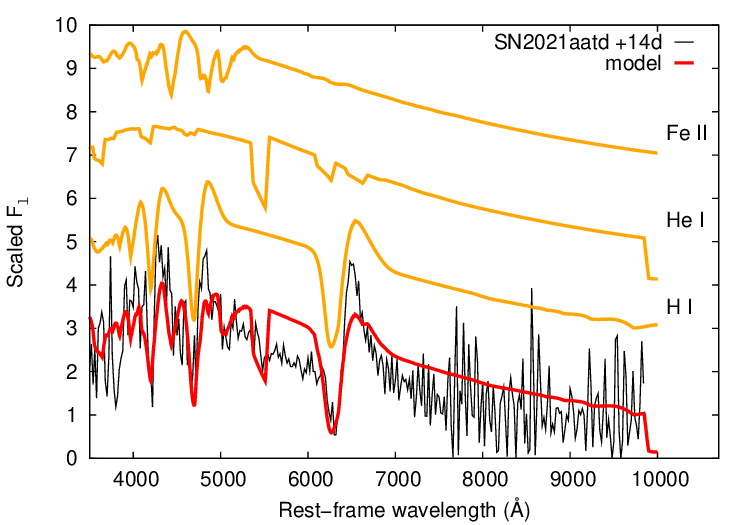}
\includegraphics[width=5.5cm]{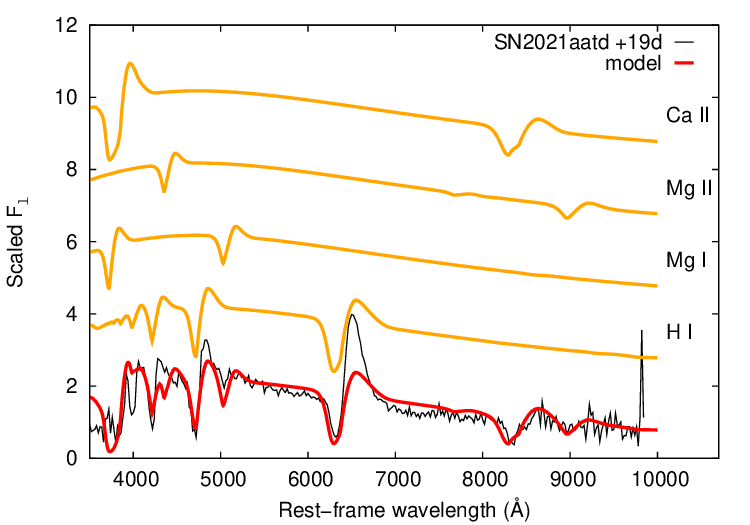}
\includegraphics[width=5.5cm]{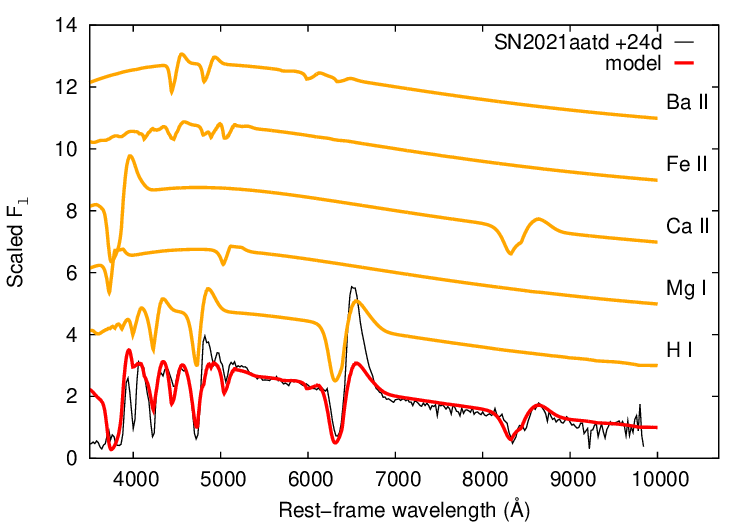}
\includegraphics[width=5.5cm]{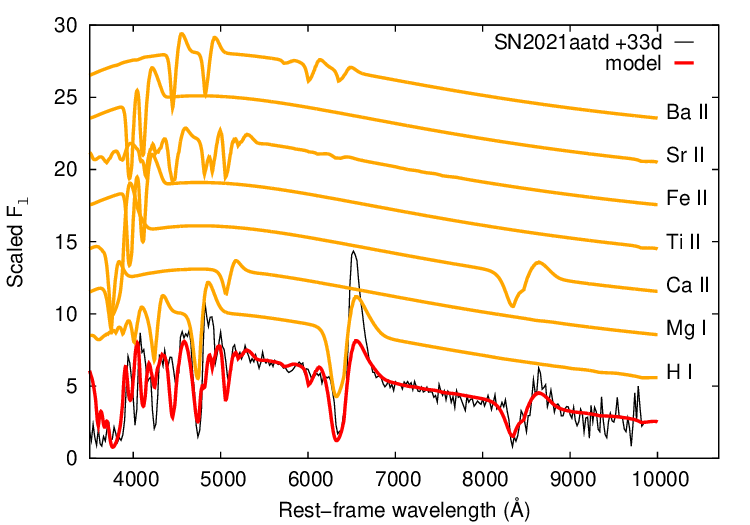}
\includegraphics[width=5.5cm]{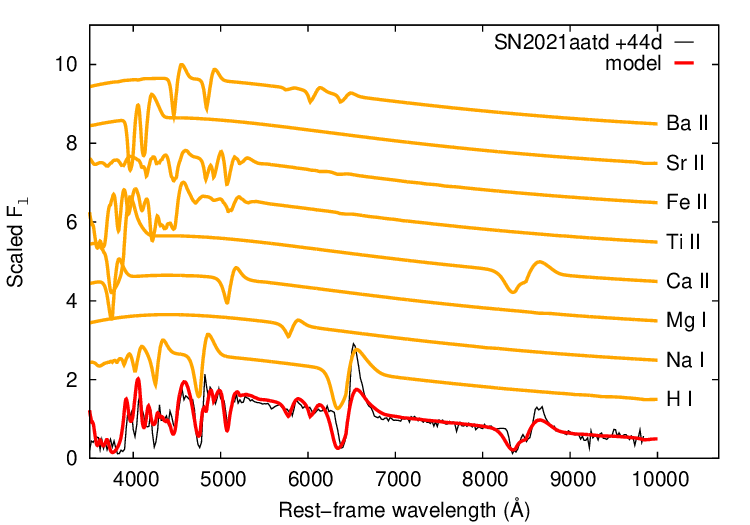}
\includegraphics[width=5.5cm]{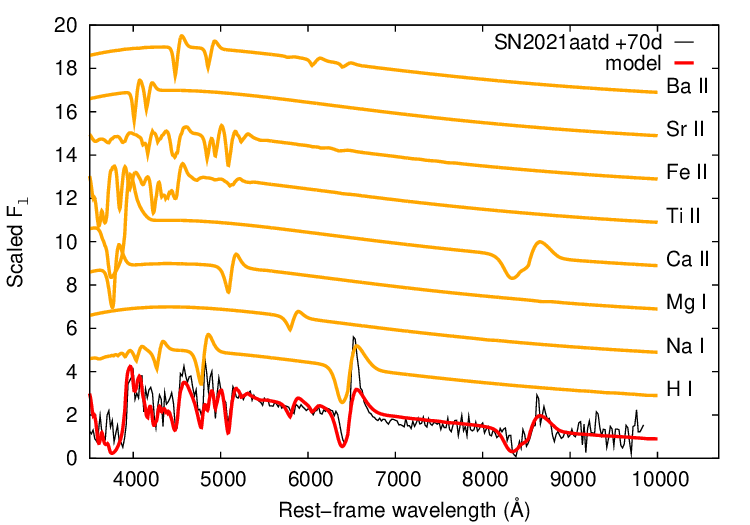}
\includegraphics[width=5.5cm]{21aatd_20211229_ions.eps}
\includegraphics[width=5.5cm]{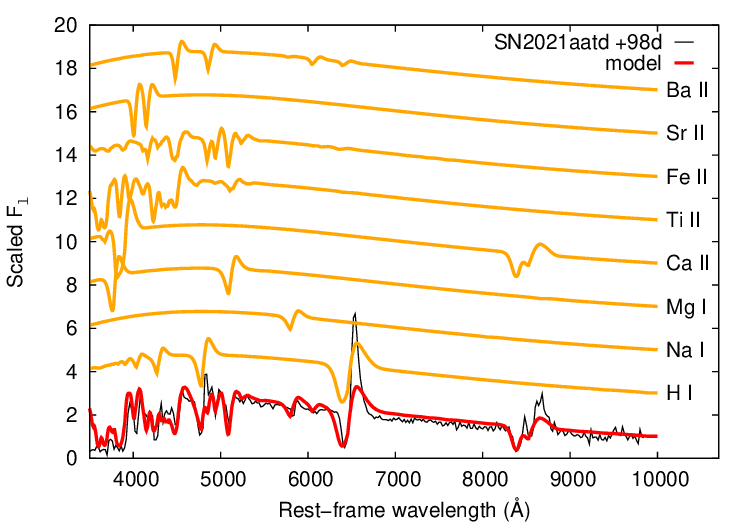}
\caption{The modeling of the available spectra of SN~2021aatd using the {\tt SYN++} code. The observed spectra (black) are corrected for redshift and interstellar reddening before plotting. Their best-fit models are marked with red color, while the vertically shifted orange lines represent the single-ion contributions to the spectra. }
\label{fig:21aatd_model}
\end{figure*}

\centering
\begin{longtable}{lcccccccccc}
\caption{Best-fit {\tt SYN++} parameter values of the modeled spectra of SN~2021aatd. The global parameters are: $T_{\rm ph}$ (1000K) and $v_{\rm ph}$ ($10^3$ km s$^{-1}$). The local parameters are: $\log\tau$ (--),  $v_{\rm min}$ ($10^3$ km s$^{-1}$), $v_{\rm max}$ ($10^3$ km s$^{-1}$), aux ($10^3$ km s$^{-1}$) and $T_{\rm exc}$ (1000 K).}
\label{tab:local_21aatd} \\
\hline
\hline
Ions & H I &  He I &  Na I &  Mg I & Mg II & Ca II & Ti II & Fe II & Sr II &  \ion{Ba}{ii}  \\
\hline
  \multicolumn{11}{c}{SN~2021aatd +2 days phase ($T_{\rm ph}~=~$9; $v_{\rm ph}~=~$16) } \\
\hline
$\log\tau$ & 1.3 & 0.0 &&&&&&&&\\
$v_{\rm min}$ & 16.0& 16.0 &&&&&&&&\\
$v_{\rm max}$ &30.0& 30.0 &&&&&&&&\\
aux & 1.0& 2.0 &&&&&&&&\\
$T_{\rm exc}$ &8.0 & 25.5 &&&&&&&&\\
\hline
  \multicolumn{11}{c}{SN~2021aatd +4 days phase ($T_{\rm ph}~=~$11; $v_{\rm ph}~=~$15) }\\
\hline
$\log\tau$ & 1.2 & -0.2 &&&&&&&&\\
$v_{\rm min}$ & 15.0& 15.0 &&&&&&&&\\
$v_{\rm max}$ &30.0& 30.0 &&&&&&&&\\
aux & 1.0& 2.0 &&&&&&&&\\
$T_{\rm exc}$ &7.5 & 7.5 &&&&&&&&\\
\hline
  \multicolumn{11}{c}{SN~2021aatd +6 days phase ($T_{\rm ph}~=~$7.5; $v_{\rm ph}~=~$14)} \\
\hline
$\log\tau$ & 1.0 &  &&&&&&&&\\
$v_{\rm min}$ & 14.0&  &&&&&&&&\\
$v_{\rm max}$ &30.0&  &&&&&&&&\\
aux & 2.0& &&&&&&&&\\
$T_{\rm exc}$ &2.6 &  &&&&&&&&\\
\hline
  \multicolumn{11}{c}{SN~2021aatd +14 days phase ($T_{\rm ph}~=~$6.5; $v_{\rm ph}~=~$10)} \\
\hline
$\log\tau$ & 1.9 &0.0  &&&&&&0.5&&\\
$v_{\rm min}$ & 10.0&20.0  &&&&&&10.0&&\\
$v_{\rm max}$ &30.0&  30.0&&&&&&30.0&&\\
aux & 2.0&10.0 &&&&&&1.0&&\\
$T_{\rm exc}$ &6.5 &  5.5&&&&&&5.5&&\\
\hline
  \multicolumn{11}{c}{SN~2021aatd +19 days phase ($T_{\rm ph}~=~$6; $v_{\rm ph}~=~$9)} \\
\hline
$\log\tau$ & 1.7 &&& 0.0 &0.0&2.5&&&&\\
$v_{\rm min}$ & 9.0&&& 9.0 &9.0&9.0&&&&\\
$v_{\rm max}$ &30.0&&& 30.0 &30.0&30.0&&&&\\
aux & 2.0&&& 2.0&2.0&2.0&&&&\\
$T_{\rm exc}$ &5.0 &&&5.0&5.0&5.5&&&&\\
\hline
  \multicolumn{11}{c}{SN~2021aatd +24 days phase ($T_{\rm ph}~=~$6; $v_{\rm ph}~=~$8)} \\
\hline
$\log\tau$ & 1.7 &&& -0.3  && 2.3&& 0.0&& 0.5\\
$v_{\rm min}$ & 8.0&&& 8.0 &&8.0 &&8.0&&8.0\\
$v_{\rm max}$ &30.0&&&  30.0&& 30.0&&30.0&&30.0\\
aux & 2.0&&& 2.0&&2.0&&1.0&&1.0\\
$T_{\rm exc}$ &5.0&&& 5.0&&5.5&&5.0&&5.0\\
\hline
  \multicolumn{11}{c}{SN~2021aatd +33 days phase ($T_{\rm ph}~=~$6; $v_{\rm ph}~=~$7)} \\
\hline
$\log\tau$ & 1.7 & & & -0.3&&2.3&-0.5 &0.6&2.2&0.9\\
$v_{\rm min}$ & 7.0  &&& 7.0&&7.0&7.0&7.0&7.0&7.0\\
$v_{\rm max}$ &30.0  &&&30.0&&30.0&30.0&30.0&30.0&30.0\\
aux & 2.0 &&&2.0&&2.0&1.0&1.0&1.0&1.0\\
$T_{\rm exc}$ & 5.0  &&&5.0&&5.5&5.0&5.0&5.0&5.0\\
\hline
  \multicolumn{11}{c}{SN~2021aatd +44 days phase ($T_{\rm ph}~=~$6.5; $v_{\rm ph}~=~$6)} \\
\hline
$\log\tau$ & 1.7 &&-0.3 & 0.0&&2.6&0.5 &0.5&2.2&0.9\\
$v_{\rm min}$ & 6.0  &&6.0& 6.0&&6.0&6.0&6.0&6.0&6.0\\
$v_{\rm max}$ &30.0  &&30.0 &30.0&&30.0&30.0&30.0&30.0&30.0\\
aux & 2.0 &&2.0&2.0&&2.0&1.0&1.0&1.0&1.0\\
$T_{\rm exc}$ & 5.0  &&5.0&5.0&&5.5&5.0&5.0&5.0&5.0\\
\hline
  \multicolumn{11}{c}{SN~2021aatd +70 days phase ($T_{\rm ph}~=~$6.5; $v_{\rm ph}~=~$5)} \\
\hline
$\log\tau$ & 1.3 &&-0.3 & 0.0&&2.9&0.1 &0.5&0.5&0.5\\
$v_{\rm min}$ & 5.0  &&5.0& 5.0&&6.0&5.0&6.0&5.0&5.0\\
$v_{\rm max}$ &30.0  &&30.0 &30.0&&30.0&30.0&30.0&30.0&30.0\\
aux & 2.0 &&2.0&2.0&&2.0&1.0&1.0&1.0&1.0\\
$T_{\rm exc}$ & 5.0  &&5.0&5.0&&5.5&5.0&5.0&5.0&5.0\\
\hline
  \multicolumn{11}{c}{SN~2021aatd +83 days phase ($T_{\rm ph}~=~$6.3; $v_{\rm ph}~=~$5)} \\
\hline
$\log\tau$ & 1.3 &&-0.3 & 0.0&&2.9&-0.3 &0.3&0.5&0.5\\
$v_{\rm min}$ & 5.0  &&5.0& 5.0&&6.0&5.0&6.0&5.0&5.0\\
$v_{\rm max}$ &30.0  &&30.0 &30.0&&30.0&30.0&30.0&30.0&30.0\\
aux & 2.0 &&2.0&2.0&&2.0&1.0&1.0&1.0&1.0\\
$T_{\rm exc}$ & 5.0  &&5.0&5.0&&5.5&5.0&5.0&5.0&5.0\\
\hline
  \multicolumn{11}{c}{SN~2021aatd +98 days phase ($T_{\rm ph}~=~$6; $v_{\rm ph}~=~$5)} \\
\hline
$\log\tau$ & 1.3 &&-0.3 & 0.0&&2.9&0.3 &0.5&1.0&0.5\\
$v_{\rm min}$ & 5.0  &&5.0& 5.0&&6.0&5.0&6.0&5.0&5.0\\
$v_{\rm max}$ &30.0  &&30.0 &30.0&&30.0&30.0&30.0&30.0&30.0\\
aux & 2.0 &&2.0&2.0&&2.0&1.0&1.0&1.0&1.0\\
$T_{\rm exc}$ & 5.0  &&5.0&5.0&&5.5&5.0&5.0&5.0&5.0\\
\hline
\end{longtable}

\newpage
\subsection{SN~2020faa}

\begin{figure*}
\centering
\includegraphics[width=5.5cm]{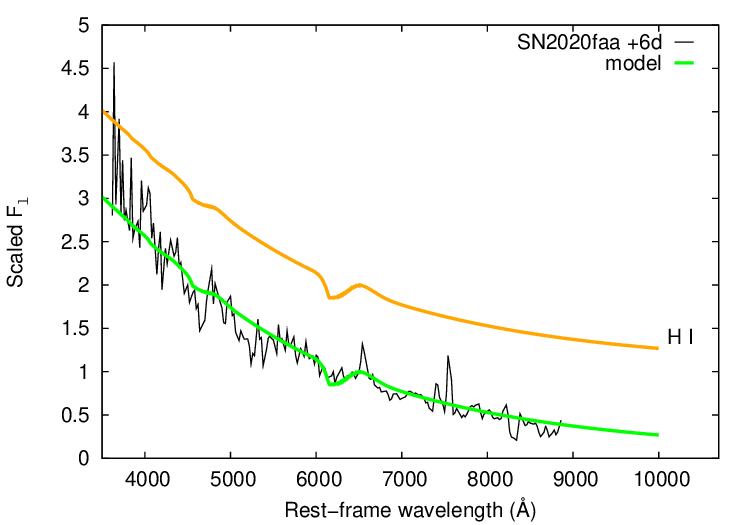}
\includegraphics[width=5.5cm]{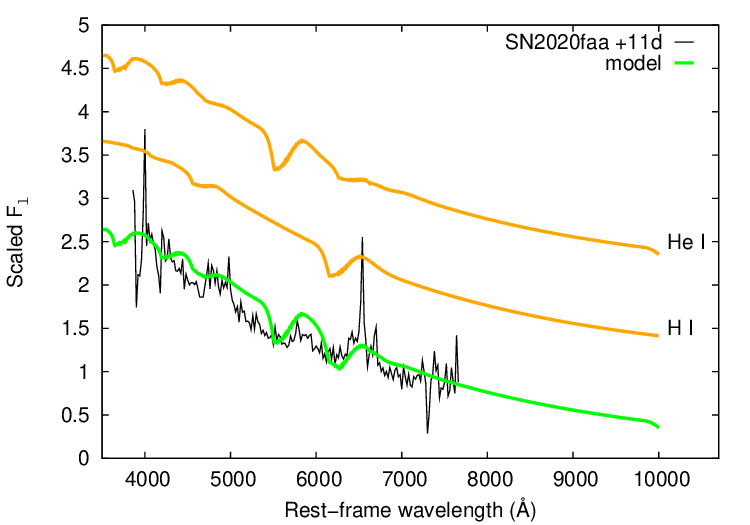}
\includegraphics[width=5.5cm]{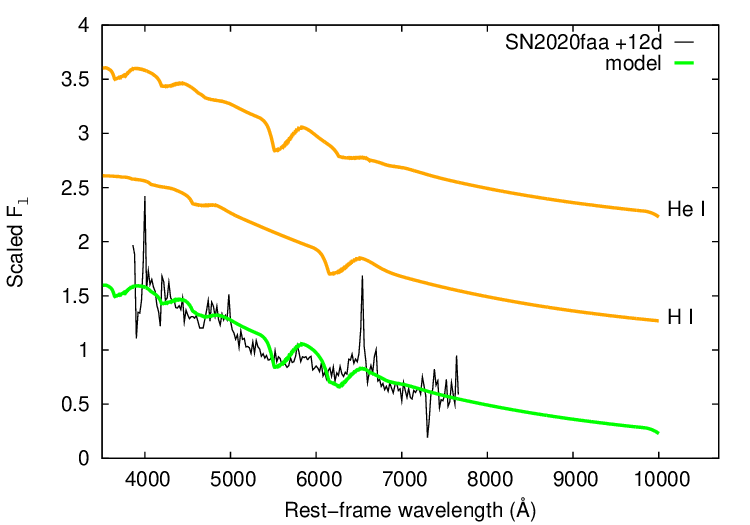}
\includegraphics[width=5.5cm]{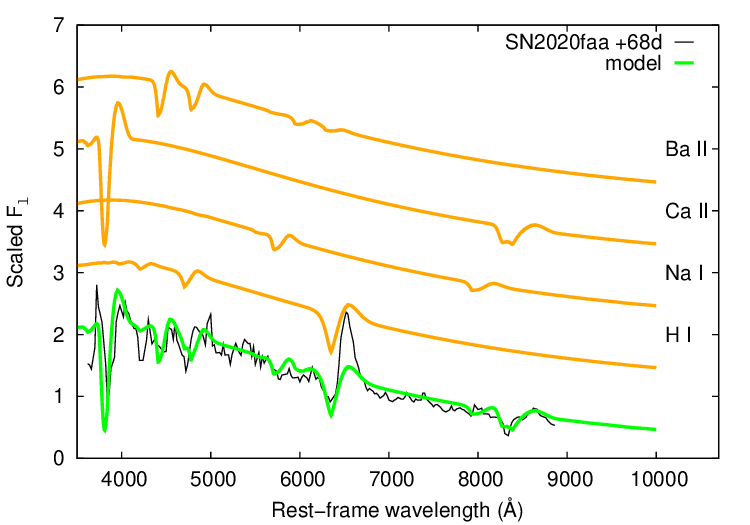}
\includegraphics[width=5.5cm]{20faa_20200621_ions.eps}
\includegraphics[width=5.5cm]{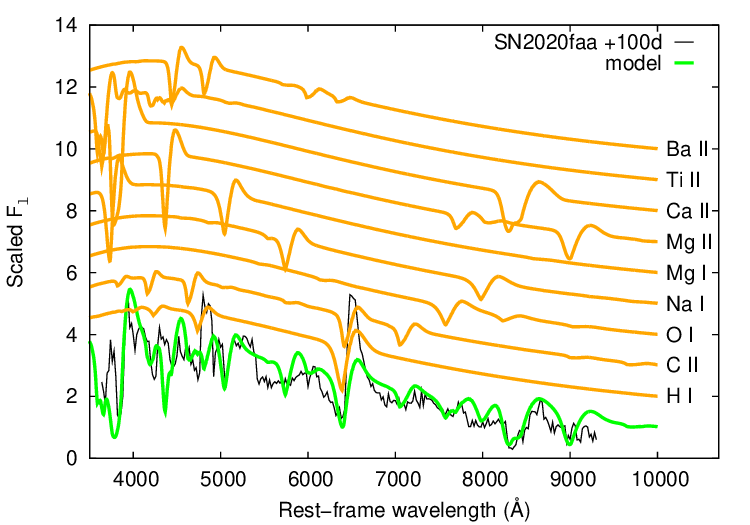}
\includegraphics[width=5.5cm]{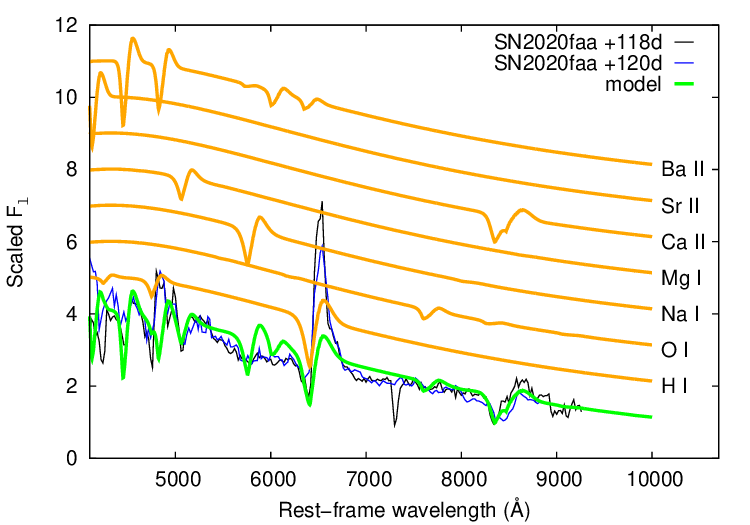}
\includegraphics[width=5.5cm]{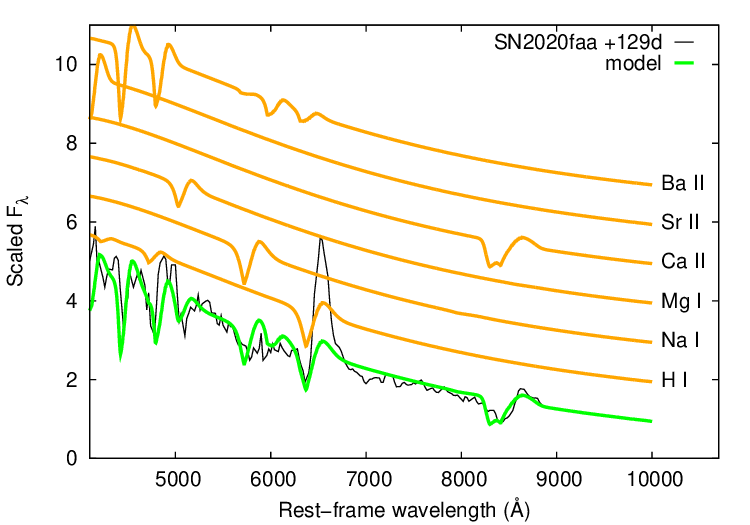}
\includegraphics[width=5.5cm]{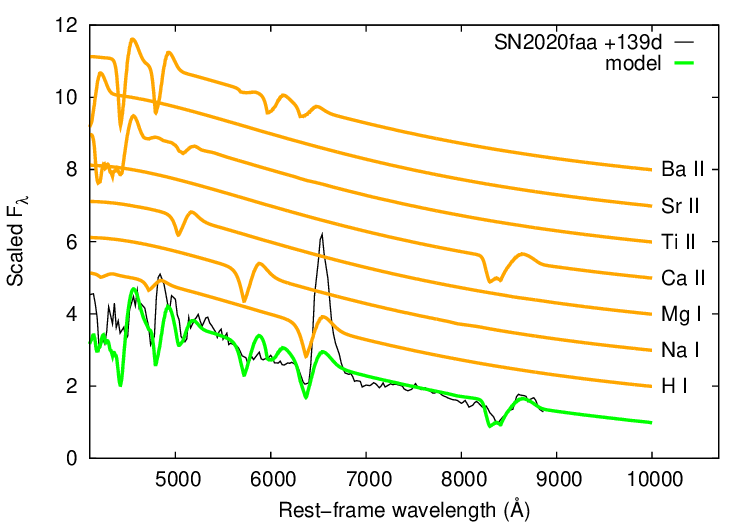}
\includegraphics[width=5.5cm]{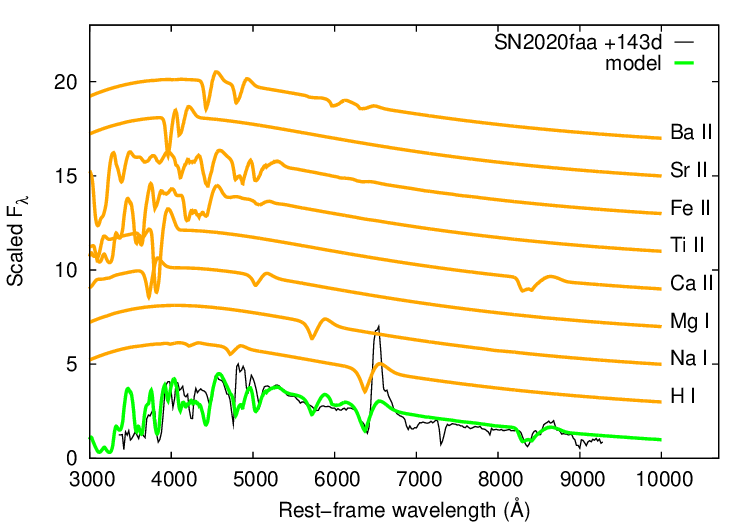}
\includegraphics[width=5.5cm]{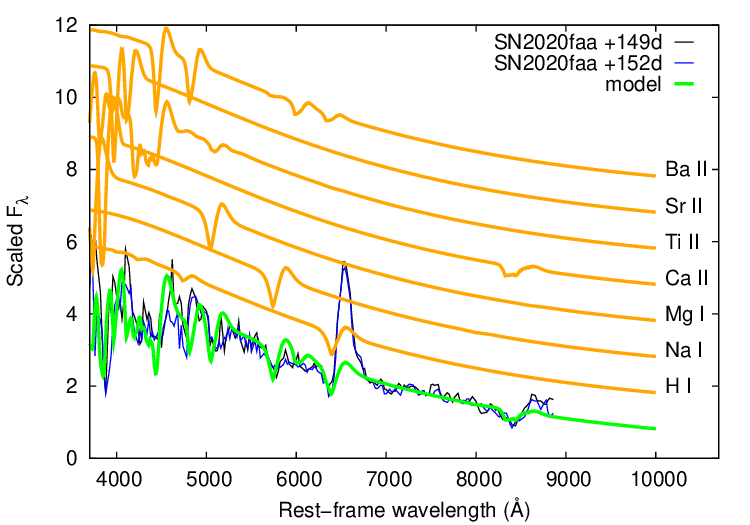}
\includegraphics[width=5.5cm]{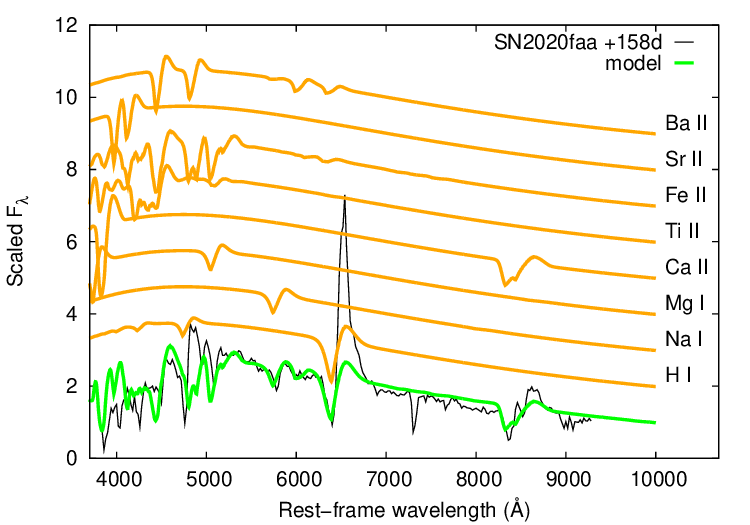}
\includegraphics[width=5.5cm]{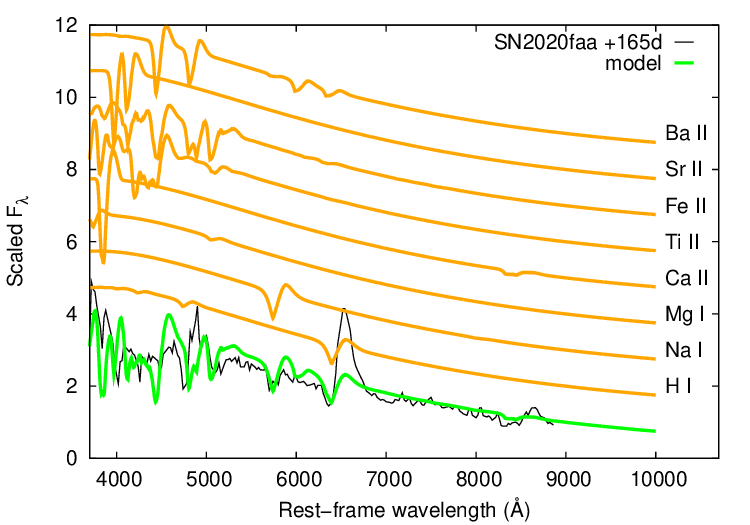}
\caption{Spectrum modeling of SN~2020faa, the comparison object of SN~2021aatd. The redshift- and reddening corrected spectra (black) are plotted together with the best-fit model obtained in {\tt SYN++} (green), and the single-ion contributions (orange) similarly to Fig. \ref{fig:21aatd_model}.}
\label{fig:20faa_model}
\end{figure*}

\centering
\begin{longtable}{lccccccccccc}
\caption{Best-fit parameters of the {\tt SYN++} modeling of SN~2020faa in the same way as in Table \ref{tab:local_21aatd}. }
\label{tab:local_20faa} \\
\hline
\hline
Ions & H I &  He I & C II& O I & Na I &  Mg I  & Ca II & Ti II & Fe II & Sr II &  \ion{Ba}{ii}  \\
\hline
  \multicolumn{12}{c}{SN~2020faa +6 days phase ($T_{\rm ph}~=~$10; $v_{\rm ph}~=~$20) } \\
\hline
$\log\tau$ & 0.3 &&&&&&&&&&\\
$v_{\rm min}$ & 20.0&&&&&&&&&&\\
$v_{\rm max}$ &30.0&&&&&&&&&&\\
aux & 2.0&&&&&&&&&&\\
$T_{\rm exc}$ &6.0 &&&&&&&&&&\\
\hline
  \multicolumn{12}{c}{SN~2020faa +11 days phase ($T_{\rm ph}~=~$8; $v_{\rm ph}~=~$20) } \\
\hline
$\log\tau$ & 0.3 &0.0&&&&&&&&&\\
$v_{\rm min}$ & 20.0&20.0&&&&&&&&&\\
$v_{\rm max}$ &30.0&30.0&&&&&&&&&\\
aux & 2.0&2.0&&&&&&&&&\\
$T_{\rm exc}$ &6.0 &6.3&&&&&&&&&\\
\hline
  \multicolumn{12}{c}{SN~2020faa +12 days phase ($T_{\rm ph}~=~$7.8; $v_{\rm ph}~=~$20) } \\
\hline
$\log\tau$ & 0.3 &0.0&&&&&&&&&\\
$v_{\rm min}$ & 20.0&20.0&&&&&&&&&\\
$v_{\rm max}$ &30.0&30.0&&&&&&&&&\\
aux & 2.0&2.2&&&&&&&&&\\
$T_{\rm exc}$ &6.0 &6.3&&&&&&&&&\\
\hline
  \multicolumn{12}{c}{SN~2020faa +68 days phase ($T_{\rm ph}~=~$7.3; $v_{\rm ph}~=~$10) } \\
\hline
$\log\tau$ & 0.5 &&&& -0.1 &&2.7 &&&&0.5\\
$v_{\rm min}$ & 10.0&&&& 10.0&&10.0 &&&&10.0\\
$v_{\rm max}$ &30.0&&&&30.0&& 30.0&&&&30.0\\
aux & 2.0&&&&1.0&& 0.7&&&&1.0\\
$T_{\rm exc}$ &5.0 &&&&20.0&&5.5&&&&5.0\\
\hline
  \multicolumn{12}{c}{SN~2020faa +88 days phase ($T_{\rm ph}~=~$7.3; $v_{\rm ph}~=~$10) } \\
\hline
$\log\tau$ & 0.0 &&&& -0.4 &0.0 &2.5 &&&&0.6\\
$v_{\rm min}$ & 10.0&&&& 10.0& 10.0&10.0 &&&&10.0\\
$v_{\rm max}$ &30.0&&&&30.0&30.0& 30.0&&&&30.0\\
aux & 2.0&&&&2.0& 1.0&2.0&&&&1.0\\
$T_{\rm exc}$ &5.0 &&&&20.0&5.0&5.5&&&&5.0\\
\hline
  \multicolumn{12}{c}{SN~2020faa +100 days phase ($T_{\rm ph}~=~$6.8; $v_{\rm ph}~=~$8) } \\
\hline
$\log\tau$ & 0.6 &&-0.3&-0.2&&0.0&3.5&-0.7&&&0.6\\
$v_{\rm min}$ & 8.0&&8.0&8.0&&8.0&8.0&8.0&&&8.0\\
$v_{\rm max}$ &30.0&&30.0&30.0&&30.0&30.0&30.0&&&30.0\\
aux & 2.0&&2.0&2.0&&1.0&1.0&1.0&&&2.0\\
$T_{\rm exc}$ &5.0 &&5.0&5.0&&5.0&5.5&5.0&&&5.0\\
\hline
  \multicolumn{12}{c}{SN~2020faa +118 and +120 days phase ($T_{\rm ph}~=~$6.6; $v_{\rm ph}~=~$7) } \\
\hline
$\log\tau$ & 0.4 &&&-0.5&-0.1&-0.5&2.3&&&1.0&0.7\\
$v_{\rm min}$ & 7.0&&&7.0&7.0&7.0&7.0&&&7.0&7.0\\
$v_{\rm max}$ &30.0&&&30.0&30.0&30.0&30.0&&&30.0&30.0\\
aux & 2.0&&&1.0&2.0&2.0&1.0&&&1.0&1.0\\
$T_{\rm exc}$ &5.0 &&&5.0&5.0&5.0&5.5&&&5.0&5.0\\
\hline
  \multicolumn{12}{c}{SN~2020faa +129 days phase ($T_{\rm ph}~=~$7.5; $v_{\rm ph}~=~$9) } \\
\hline
$\log\tau$ & 0.2 &&&&-0.1 &-0.5&3.0&&&1.0&0.8\\
$v_{\rm min}$ & 9.0&&&&9.0&9.0&9.0&&&9.0&9.0\\
$v_{\rm max}$ &30.0&&&&30.0&30.0&30.0&&&30.0&30.0\\
aux & 2.0&&&&2.0&2.0&0.6&&&1.0&1.0\\
$T_{\rm exc}$ &5.0 &&&&5.0&5.0&5.5&&&5.0&5.0\\
\hline
  \multicolumn{12}{c}{SN~2020faa +139 days phase ($T_{\rm ph}~=~$7; $v_{\rm ph}~=~$9) } \\
\hline
$\log\tau$ & 0.2 &&&&-0.1&-0.5&3.0&0.0&&1.0&0.9\\
$v_{\rm min}$ & 9.0&&&&9.0&9.0&9.0&9.0&&9.0&9.0\\
$v_{\rm max}$ &30.0&&&&30.0&30.0&30.0&30.0&&30.0&30.0\\
aux & 2.0&&&&2.0&2.0&0.6&2.0&&1.0&1.0\\
$T_{\rm exc}$ &5.0 &&&&5.0&5.0&5.5&5.0&&5.0&5.0\\
\hline
  \multicolumn{12}{c}{SN~2020faa +143 days phase ($T_{\rm ph}~=~$7.2; $v_{\rm ph}~=~$9) } \\
\hline
$\log\tau$ & 0.4 &&&&-0.1&-0.5&3.0&0.0&0.6&1.0&0.7\\
$v_{\rm min}$ & 9.0&&&&9.0&9.0&9.0&9.0&9.0&9.0&9.0\\
$v_{\rm max}$ &30.0&&&&30.0&30.0&30.0&30.0&30.0&30.0&30.0\\
aux & 2.0&&&&2.0&2.0&0.6&1.0&1.0&1.0&1.0\\
$T_{\rm exc}$ &5.0 &&&&5.0&5.0&5.5&5.0&5.0&5.0&5.0\\
\hline
  \multicolumn{12}{c}{SN~2020faa +149 and +152 days phase ($T_{\rm ph}~=~$8.0; $v_{\rm ph}~=~$8) } \\
\hline
$\log\tau$ & 0.0 &&&&-0.1&-0.2&2.2&0.0&&1.0&0.7\\
$v_{\rm min}$ & 8.0&&&&8.0&8.0&8.0&8.0&&8.0&8.0\\
$v_{\rm max}$ &30.0&&&&30.0&30.0&30.0&30.0&&30.0&30.0\\
aux & 2.0&&&&2.0&2.0&0.6&2.0&&1.0&1.0\\
$T_{\rm exc}$ &5.0 &&&&5.0&5.0&5.5&5.0&&5.0&5.0\\
\hline
  \multicolumn{12}{c}{SN~2020faa +158 days phase ($T_{\rm ph}~=~$6.0; $v_{\rm ph}~=~$8) } \\
\hline
$\log\tau$ & 0.5 &&&&-0.3&-0.4&2.9&0.0&0.8&1.0&0.7\\
$v_{\rm min}$ & 8.0&&&&8.0&8.0&8.0&8.0&8.0&8.0&8.0\\
$v_{\rm max}$ &30.0&&&&30.0&30.0&30.0&30.0&30.0&30.0&30.0\\
aux & 2.0&&&&2.0&2.0&0.6&1.0&1.0&1.0&1.0\\
$T_{\rm exc}$ &5.0 &&&&5.0&5.0&5.5&5.0&5.0&5.0&5.0\\
\hline
  \multicolumn{12}{c}{SN~2020faa +165 days phase ($T_{\rm ph}~=~$7.5; $v_{\rm ph}~=~$8) } \\
\hline
$\log\tau$ & 0.0 &&&&-0.1&-1.3&1.7&0.0&0.5&1.0&0.7\\
$v_{\rm min}$ & 8.0&&&&8.0&8.0&8.0&8.0&8.0&8.0&8.0\\
$v_{\rm max}$ &30.0&&&&30.0&30.0&30.0&30.0&30.0&30.0&30.0\\
aux & 2.0&&&&2.0&2.0&0.6&1.0&1.0&1.0&1.0\\
$T_{\rm exc}$ &5.0 &&&&5.0&5.0&5.5&5.0&5.0&5.0&5.0\\
\hline
\end{longtable}

\newpage
\subsection{SN~1987A}
%
%
\onecolumn
\centering
\begin{longtable}{lcccccccccc}
\caption{Best-fit parameters of the {\tt SYN++} modeling of SN~1987A in the same way as in Table \ref{tab:local_21aatd}. }
\label{tab:local_87a} \\
\hline
\hline
Ions & H I & N II & O I & Na I &  Mg II  & Ca II & Ti II & Fe II & Sr II &  \ion{Ba}{ii}  \\
\hline
  \multicolumn{11}{c}{SN~1987A +84 days phase ($T_{\rm ph}~=~$4.5; $v_{\rm ph}~=~$3) } \\
\hline
$\log\tau$ & 1.7 & 0.0 &1.0&1.3&-0.3&2.3&-0.5&1.8&2.2&1.2\\
$v_{\rm min}$ & 3.0&3.0&3.0&3.0&3.0&3.0&3.0&3.0&3.0&3.0\\
$v_{\rm max}$ &30.0&30.0&30.0&30.0&30.0&30.0&30.0&30.0&30.0&30.0\\
aux & 1.0&1.0&1.0&1.0&2.0&2.0&1.0&0.5&1.0&1.0\\
$T_{\rm exc}$ &5.0&14.5&5.0&5.0&5.0&5.5&5.0&5.0&5.0&4.5\\
\hline
\end{longtable}

%
%
%
\end{appendix}

\end{document}